%% file: main.tex
\documentclass[journal]{IEEEtran}
\usepackage{amsmath,amsfonts}
\usepackage{algorithmic}
\usepackage{algorithm}
\usepackage{array}
\usepackage[caption=false,font=normalsize,labelfont=sf,textfont=sf]{subfig}
\usepackage{textcomp}
\usepackage{stfloats}
\usepackage{url}
\usepackage{verbatim}
\usepackage{graphicx}
\usepackage{cite}
\usepackage{amsmath,amsfonts,amssymb}
\usepackage{bm}
\usepackage{algorithmic}
\usepackage{algorithm}
\usepackage{array}
\usepackage{booktabs}
\usepackage{multirow}
\usepackage[table]{xcolor}
\usepackage{hyperref}
\usepackage{balance}
\usepackage{amsthm}
\usepackage{marvosym}
\usepackage{tabularx}
\usepackage{makecell}
\usepackage{ragged2e}
\usepackage[flushleft]{threeparttable}

\usepackage{tikz}
\usetikzlibrary{positioning,arrows.meta,fit,backgrounds,decorations.pathreplacing}

\usepackage[most]{tcolorbox}
\newtcolorbox{theorybox}{
colback=gray!5,
colframe=black,
boxrule=0.8pt,
arc=2pt,
left=6pt,
right=6pt,
top=6pt,
bottom=6pt
}

\newtcolorbox{findingbox}[1]{colback=gray!10!white,
  colframe=gray!70!black,
  coltitle=white,
  boxrule=0.8pt,
  arc=2mm,
  left=2mm,
  right=2mm,
  top=1mm,
  bottom=1mm,
  title=Findings (#1)}

\graphicspath{{images/}}

\hypersetup{
    colorlinks=true,
    linkcolor=blue,
    citecolor=blue,
    urlcolor=blue
}

\newcommand{\rqone}{\textbf{RQ1:} \textit{Do LLM4Code models consistently achieve better performance when using Chain-of-Thought (CoT)?}}
\newcommand{\rqtwo}{\textbf{RQ2:} \textit{Does CoT improve the robustness of LLM4Code models under perturbed prompts?}}
\newcommand{\rqthree}{\textbf{RQ3:} \textit{Is early-stage uncertainty in the generation process predictive of final code-generation failure?}}
\newcommand{\rqfour}{\textbf{RQ4:} \textit{How do input perturbations affect the structure of CoT reasoning trajectories?}}

\newcommand{\heng}[1]{}
\newcommand{\armstrong}[1]{}
\newcommand{\da}[1]{}
\newcommand{\Foutse}[1]{}
\newcommand{\accepted}[1]{}

\begin{document}

\title{Structural Anchors and Reasoning Fragility: Understanding CoT Robustness in LLM4Code}

\author{\IEEEauthorblockN{Yang Liu, Da Song $^{\textrm{\Letter}}$, 
         Armstrong Foundjem,~\IEEEmembership{SMIEEE, }
         Heng Li,~\IEEEmembership{MIEEE, } Foutse Khomh,~\IEEEmembership{SMIEEE}
         }
\IEEEcompsocitemizethanks{
        \IEEEcompsocthanksitem $^{\textrm{\Letter}}$ Corresponding author
        \IEEEcompsocthanksitem Yang Liu, Polytechnique Montreal, CA. Yang-2.liu@polymtl.ca
        \IEEEcompsocthanksitem Da Song, School of Cryptologic Science and Engineering, Shandong University, Jinan, Shandong, China, song\_da@sdu.edu.cn
        \IEEEcompsocthanksitem Armstrong~F., Polytechnique~Montreal, CA. foundjem@ieee.org
        \IEEEcompsocthanksitem Heng Li, Polytechnique Montreal, CA. heng.li@polymtl.ca
        \IEEEcompsocthanksitem Foutse K., Polytechnique Montreal, CA. foutse.khomh@polymtl.ca
         
    }
}



\maketitle

\begin{abstract}
\textbf{Context.} Chain-of-Thought (CoT) prompting is widely used to elicit explicit reasoning from large language models for code (LLM4Code). However, its impact on robustness and the stability of reasoning trajectories under realistic input perturbations remains poorly understood. Prior work has largely evaluated CoT through final correctness, leaving a critical gap in our understanding of how CoT reshapes internal uncertainty dynamics and why it sometimes harms rather than helps code generation. In this paper, it suggests that CoT is not uniformly beneficial for LLM4Code; instead, its robustness may depend on whether perturbations destabilize structurally sensitive commitment points along the reasoning-to-code trajectory. \textbf{Approach.} We conduct a controlled, large-scale empirical study of CoT across six models and two challenging code benchmarks (MHPP and BigCodeBench), subjecting task docstrings to systematic character-, word-, and sentence-level perturbations. We instrument full generation traces with token-level uncertainty (entropy and probability differential), 
define three novel \emph{structural anchors} along the reasoning--code trajectory (reasoning--code transition, symbolic commitment, and algorithmic articulation), and analyze how perturbations deform CoT trajectories relative to these anchors. \textbf{Findings.} (1) CoT does \emph{not} yield uniform performance or robustness gains: its benefits are contingent on model family, task structure, and prompt explicitness. (2) CoT and No-CoT exhibit \emph{distinct robustness profiles} rather than a simple dominance ordering, with different perturbation families triggering different failure modes. (3) We identify three recurrent trajectory deformations—\emph{Lengthening, Branching, and Simplification}—that systematically emerge when perturbations interact with structural anchors and explain differential success/failure patterns. (4) Early-stage uncertainty is only weakly predictive of final correctness (AUROC $\approx$ 0.55--0.60), but it serves as a reliable \emph{diagnostic signal} for localizing where trajectory instability begins around sensitive anchors. \textbf{Implications.} These results provide a unified explanation for CoT’s mixed empirical performance in code generation and suggest concrete design principles—anchor-aware training, perturbation-aware prompting, and uncertainty-guided trajectory monitoring—for building more robust reasoning-based code generators. 
\emph{Critical takeaway:} \textbf{CoT is not universally beneficial for code generation robustness; its value depends on how perturbations interact with structurally sensitive anchors in the reasoning-to-code trajectory. When these anchors are destabilized, reasoning trajectories frequently exhibit lengthening, branching, or simplification patterns that correlate with downstream generation failures.}
\end{abstract}


\input{introduction}
\input{Background}
\input{Relatedwork}
\input{Methodolog}
\input{Results}
\input{Discuss}
\input{Implication}
\input{LimitationAndFutureWork}
\input{ThreatstoValidity}
\input{Conclusions}
\input{Acknowledgement}

\noindent\textbf{Data Availability Statement} The replication package, including our datasets and more results, is available publicly at: 
\url{https://surl.li/nubody}

\bibliographystyle{IEEEtran}
\bibliography{duplicate}


 





\end{document}

%% file: Introduction.tex
\section{Introduction}\label{sec:intro}
Large language models for code (LLM4Code) have recently achieved remarkable progress in translating natural language descriptions into executable programs. Models such as CodeLlama~\cite{roziere2023code}, DeepSeek-Coder~\cite{guo2024deepseek}, and Qwen~\cite{hui2024qwen2} demonstrate impressive capabilities across diverse programming tasks, yet their performance remains unstable under slight variations in input prompts~\cite{liu2025adversarial, wang2022recode, yang2024important, mastropaolo2023robustness, zhou2022adversarial}. Even minor linguistic perturbations or paraphrases can lead to drastic differences in reasoning behavior and output correctness~\cite{yan2025robustness}. As these models are increasingly integrated into software development workflows, understanding the sources and dynamics of such instability has become a pressing research concern~\cite{zhong2024can}.

Beyond standalone code synthesis, LLM4Code models are increasingly being embedded into agentic software engineering systems that iteratively plan, generate, test, revise, and interact with external tools~\cite{yang2024swe, liu2024large}. In such settings, robustness is no longer only about whether a model produces a correct final program, but also about whether its intermediate reasoning remains stable under natural variation in user instructions, contextual noise, or adversarially perturbed inputs. A small perturbation at the prompt level may not merely change a single output token; it can alter the entire downstream reasoning trajectory, misguide tool use, and amplify error propagation across multiple decision steps. This makes robust reasoning a central requirement for the safe and reliable deployment of LLM-based coding agents.

A prominent line of research advocates chain-of-thought (CoT) prompting~\cite{wei2022chain} as a means to enhance model reasoning by decomposing a task into intermediate logical steps. While CoT has been shown to improve accuracy and interpretability in complex tasks ~\cite{kojima2022large, wang2022self}, it also alters the internal reasoning trajectory of the model—often lengthening the generation process and introducing higher variability in decision paths~\cite{dhuliawala2024chain, huang2023codecot}. These changes raise an open question: \textit{how does CoT affect the model’s uncertainty during code generation, and to what extent does this uncertainty correlate with robustness or failure?}


Existing studies have largely evaluated CoT in LLM4Code through final correctness metrics such as pass@k, showing that CoT can improve performance on some reasoning-intensive tasks. However, prior work offers limited insight into \emph{how} CoT reshapes the internal reasoning process under realistic input perturbations, and whether changes in \emph{generation-time uncertainty} explain why CoT sometimes helps and sometimes harms code generation. In particular, the literature provides little systematic evidence on (i) how uncertainty evolves along the reasoning--code trajectory under CoT, (ii) whether early-stage uncertainty can reliably signal downstream failure, and (iii) how prompt perturbations interact with key structural transitions during generation. This gap is critical because LLM4Code models are increasingly deployed in safety- and reliability-sensitive software engineering workflows, where understanding the sources of instability—rather than only measuring end-to-end accuracy—is essential for building robust systems.

This challenge is especially important because reasoning has become a core mechanism through which modern LLMs achieve strong coding performance. Yet reasoning also introduces new failure surfaces~\cite{zhou2024can}: longer trajectories, more branching opportunities, delayed commitment, and greater sensitivity to early deviations. Despite growing interest in CoT for code generation, the field still lacks a process-level account of how reasoning becomes unstable, where such instability concentrates along the generation trajectory~\cite{yin2024reasoning}, and whether it can be detected before failure fully materializes. Uncertainty signals offer a promising lens for this purpose, as they can expose hesitation, ambiguity, or commitment shifts during generation and may therefore support early diagnosis, adaptive intervention, and more robust reasoning-aware system design~\cite{yin2024reasoning}.

To address this gap, we frame our investigation around four research questions (RQs):\\
\rqone\\
\rqtwo\\
\rqthree\\
\rqfour

\noindent\textbf{Contributions. }This paper makes four contributions: \textbf{(1) A Structural Theory of CoT in Code Generation.} We propose an anchor-deformation perspective that conceptualizes CoT as a trajectory-level transformation introducing structurally sensitive commitment points (``anchors'') whose stability governs correctness and robustness. \textbf{(2) A Hypothesis-Driven Empirical Evaluation.} We systematically compare CoT and No-CoT across multiple models, datasets, and perturbation families, showing that CoT’s effects are contingent rather than uniformly beneficial. \textbf{(3) An Anchor-Aligned Trajectory Analysis Framework.} We introduce spike localization and deformation categorization (lengthening, branching, simplifying) to analyze reasoning instability under perturbation. \textbf{(4) A Security-Oriented Robustness Extension.} We demonstrate that CoT exhibits semantic resilience under structural perturbations but lexical fragility under low-level noise, revealing a utility–security trade-off. 
In particular, we interpret these findings through an Anchor–Deformation perspective, which conceptualizes CoT reasoning as a trajectory structured around commitment points whose stability determines robustness and correctness.

The remainder of this paper is organized as follows. 
Section~\ref{sec:background} discusses the background of this study, and Section~\ref{sec:related_work} reviews related work.  Section~\ref{sec:methodology} presents the experimental methodology, evaluation setup, and the statistical hypotheses that guide our analysis. Section~\ref{sec:results} reports the empirical results. 
Section~\ref{sec:discussion} discusses the implications of our findings. Finally, Section~\ref{sec:conclusions} concludes the paper and outlines directions for future work.

%% file: Background.tex
\section{Background} \label{sec:background}
Recent advances in large language models for code (LLM4Code) have transformed how developers approach programming tasks~\cite{hou2024large}, enabling automatic code synthesis~\cite{chen2021evaluating}, refactoring\cite{cordeiro2024empirical}, and problem solving from natural language descriptions~\cite{austin2021program, hendrycks2021measuring}. Yet the underlying reasoning mechanisms that guide these models remain opaque and unstable. 
Understanding how these models reason and how reliable that reasoning is under varying conditions has therefore become an important research challenge.
To ground our investigation into the relationship between reasoning and uncertainty, this section reviews four interrelated areas: the development and characteristics of LLM4Code models, the emergence of CoT reasoning in program synthesis, the role of uncertainty in generation processes, and current approaches for evaluating robustness under prompt perturbations.


\subsection{A Motivating Example}

Table~\ref{tab:motivating_example_realcase} presents a real pair of generations for the same task, \texttt{MHPP/79} (\texttt{rank\_task}), produced by Qwen2.5-Coder-7B-Instruct under CoT prompting. The original prompt and the perturbed prompt describe the same scheduling problem, but the model follows different reasoning paths and produces different code structures.

\begin{table*}[t]
\centering
\footnotesize
\setlength{\tabcolsep}{4pt}
\renewcommand{\arraystretch}{1.2}
\caption{A real example from \texttt{MHPP} (\texttt{rank\_task}) comparing the original prompt and the perturbed prompt.}
\label{tab:motivating_example_realcase}

\begin{tabularx}{\textwidth}{>{\RaggedRight\arraybackslash}p{1.7cm} >{\RaggedRight\arraybackslash}X >{\RaggedRight\arraybackslash}X}
\toprule
\textbf{Part} & \textbf{Original prompt} & \textbf{Perturbed prompt} \\
\midrule

Prompt
&
\makecell[l]{\ttfamily
def rank\_task(tasks: List[List[int]]) -> List[int]:\\
"Given a list of tasks ... where\\
$tasks[i] = [start\_time, process\_time]$ ...\\
if there are multiple tasks piled up, choose\\
the task with the shortest $process\_time$ first;\\
if tied, choose the smallest index ..."\\
Example outputs: [0,2,1], [1,4,3,2,0]
}
&
\makecell[l]{\ttfamily
def rank\_task(tasks: List[List[int]]) -> List[int]:\\
Same task semantics, but the natural-language\\
description contains character-level $C_1$\\
perturbations (e.g., irregular capitalization),\\
while preserving the same scheduling rules\\
and example behavior.
}
\\

\midrule

CoT process
&
\makecell[l]{\ttfamily
1. Initialize result.\\
2. Create a priority queue pq ordered by\\
process time and index.\\
3. Push each task into pq.\\
4. While pq is not empty:\\
\quad pop the shortest task and append\\
\quad its index to result.
}
&
\makecell[l]{\ttfamily
1. Initialize result and current\_time.\\
2. While tasks remain:\\
\quad filter tasks with $start\_time \leq current\_time$;\\
\quad sort available tasks by $process\_time$ and index;\\
\quad select one task and update current\_time.
}
\\

\midrule

Code
&
\makecell[l]{\ttfamily
result = []\\
pq = []\\
for i, (start\_time, process\_time) in enumerate(tasks):\\
\quad heapq.heappush(pq, (process\_time, i))\\
while pq:\\
\quad \textunderscore, index = heapq.heappop(pq)\\
\quad result.append(index)\\
return result
}
&
\makecell[l]{\ttfamily
result = []\\
current\_time = 0\\
while tasks:\\
\quad available\_tasks = [task for task in tasks\\
\quad\quad if task[0] <= current\_time]\\
\quad available\_tasks.sort(\\
\quad\quad key=lambda x: (x[1], tasks.index(x)))\\
\quad if available\_tasks:\\
\qquad selected\_task = available\_tasks.pop(0)\\
\qquad result.append(tasks.index(selected\_task))\\
\qquad current\_time += selected\_task[1]\\
return result
}
\\

\bottomrule
\end{tabularx}
\end{table*}

This real example shows that even when task semantics remain unchanged, a small perturbation can redirect the model toward a different reasoning trajectory. In the original case, the model commits early to a compact priority-queue strategy. In the perturbed case, it instead adopts a longer simulation-style path built around \texttt{current\_time} and \texttt{available\_tasks}. The perturbation therefore changes not only the final implementation pattern, but also the intermediate reasoning process through which the code is constructed.

\subsection{LLM4Code for Code Generation}

LLM4Code models such as CodeLlama, Qwen, and DeepSeek-Coder are trained on large-scale code corpora and have demonstrated strong capabilities in program synthesis, code completion, and bug fixing~\cite{roziere2023code, guo2024deepseek, hui2024qwen2}. Like general autoregressive language models, they generate source code token by token conditioned on natural language instructions and partial code context~\cite{radford2019language}. In code generation, however, this autoregressive process is especially sensitive because local token decisions are tightly constrained by syntax, semantics, and long-range program logic. As a result, small variations in prompt phrasing, formatting, or comments can shift the generation trajectory and produce substantially different programs~\cite{liu2025adversarial, wang2022recode, yang2024important, mastropaolo2023robustness, zhou2022adversarial}. This prompt sensitivity motivates a closer examination of not only what output is produced, but also how the model arrives there.

\subsection{Chain-of-Thought Reasoning in Code Generation}


CoT prompting encourages a model to generate intermediate reasoning steps before producing the final answer~\cite{wei2022chain}. In code generation, these intermediate steps may include decomposing the task, identifying algorithmic constraints, reasoning about corner cases, or outlining solution logic before emitting executable code~\cite{li2025structured, yu2023towards}. Compared with direct generation, CoT therefore does more than add explanatory text: it reshapes the generation trajectory itself by lengthening the token sequence, introducing additional intermediate commitments, and creating more opportunities for local deviations to influence downstream code. This makes CoT potentially helpful for difficult tasks, but also potentially more vulnerable to error accumulation and instability under perturbed inputs~\cite{li2025think}. In our study, CoT is thus treated not only as a prompting strategy, but also as a trajectory-level transformation of the code-generation process.

\subsection{Uncertainty in LLM4Code}


Uncertainty characterizes how decisively a model selects the next token during generation~\cite{kuhn2023semantic, shorinwa2025survey}. In LLM4Code, this signal is particularly informative because code generation requires not only fluent continuation but also strict syntactic validity and logical consistency. Common token-level measures such as entropy and probability differential quantify, respectively, how dispersed the predictive distribution is and how sharply the model prefers one token over competing alternatives~\cite{zhu2025uncertainty}. From a process perspective, elevated uncertainty may reflect hesitation, ambiguity, or unstable reasoning commitments during generation. In code generation, such instability is important because an uncertain decision at an early stage may propagate through subsequent reasoning or code tokens and eventually lead to compilation or functional failure. For this reason, uncertainty provides a useful lens for examining how reasoning unfolds over time, rather than evaluating correctness only at the final output.

\subsection{Robustness under Prompt Perturbations}


Robustness in code generation refers to a model’s ability to preserve correct and consistent behavior under inputs that remain semantically equivalent but differ in surface form~\cite{ribeiro2020beyond, zhu2023promptrobust}. In practice, such variation may arise from paraphrasing, formatting changes, character-level noise, or alternative phrasings of the same programming task. For LLM4Code, robustness matters because even small prompt changes can alter intermediate reasoning, shift implementation choices, and ultimately affect program correctness. Most existing evaluations quantify this effect through changes in end-point metrics such as pass@k~\cite{chen2021evaluating}. While useful, such outcome-level measures cannot reveal whether failures originate from early reasoning instability, local uncertainty spikes, or gradual divergence across the generation trajectory. This motivates a process-level robustness view that jointly considers perturbation, uncertainty, and reasoning dynamics.

%% file: Relatedwork.tex
\section{Related Works} \label{sec:related_work}
In this section, we discuss three areas of research relevant to our study: the development of large language models for code generation, analyses of reasoning and CoT prompting, and uncertainty-based robustness evaluation. We review each stream and highlight how our work advances the state of the art by linking reasoning structure, uncertainty dynamics, and robustness under perturbed conditions.

\subsection{Large Language Models for Code Generation (LLM4Code)}
Recent years have witnessed rapid advances in LLM4Code.
Early pre-trained encoders such as CodeBERT~\cite{feng2020codebert} and GraphCodeBERT~\cite{guo2020graphcodebert} captured structural representations of source code but were limited by context length and bidirectional masking.
Autoregressive models such as CodeGen~\cite{nijkamp2022codegen}, CodeLlama~\cite{roziere2023code}, Qwen~\cite{bai2023qwen}, and DeepSeek-Coder~\cite{guo2024deepseek} expanded training corpora to hundreds of billions of tokens and achieved state-of-the-art results in program synthesis and repair.
These GPT-style models demonstrate strong generalization but remain highly sensitive to prompt phrasing~\cite{mastropaolo2023robustness}, causing large output variance even for semantically equivalent inputs.

Most prior studies on LLM4Code focus on improving accuracy or instruction-following capabilities rather than understanding the internal reasoning process.
Our work differs by analyzing how reasoning modes (CoT vs. direct generation) reshape internal uncertainty patterns, providing an interpretable view of why model performance fluctuates under minor input variations.

\subsection{Reasoning and CoT Prompting}
CoT prompting~\cite{wei2022chain} has been proposed to enhance the interpretability of LLM reasoning by eliciting intermediate reasoning steps that shape the model’s internal reasoning process before producing the final output.

Follow-up studies, such as Zelikman et al. ~\cite{zelikman2022star} and Yao et al. ~\cite{yao2023tree}, show that CoT can improve performance in multi-step reasoning and algorithmic tasks.
In the context of code generation, prior works such as Chain of Code~\cite{li2023chain} and Uncertainty-Guided CoT for Code Generation~\cite{zhu2025uncertainty} have mainly evaluated CoT for correctness or interpretability but seldom examined its stability or robustness.

Unlike existing work, which investigates how CoT improves model performance, our study explicitly investigates its trade-offs—quantifying how CoT affects uncertainty trajectories and robustness under semantically equivalent but perturbed prompts.
To our knowledge, this is the first large-scale analysis linking CoT reasoning to quantitative uncertainty metrics in LLM4Code.

\subsection{Uncertainty Estimation and Robustness Evaluation}
Uncertainty estimation has long been used to diagnose model reliability in NLP.
Methods based on token-level entropy, logit variance, or probability differentials~\cite{kadavath2022language} have been shown to detect low-confidence or hallucinated outputs.
In LLM4Code, several studies explore entropy as a proxy for output confidence~\cite{zhu2025uncertainty}, but these approaches often measure uncertainty post hoc—after generation—rather than as a temporal signal within reasoning.
Similarly, research on prompt robustness~\cite{liu2025adversarial} and adversarial evaluation~\cite{jha2023codeattack} focuses on final accuracy degradation, overlooking the internal dynamics that lead to instability.

Our work bridges these threads by introducing a unified evaluation framework that connects early-stage uncertainty trajectories to downstream robustness outcomes.
We quantify how CoT reasoning amplifies or suppresses uncertainty during generation and test whether these signals serve as early-warning indicators of robustness degradation under multi-level prompt perturbations.

\subsection{Comparative Positioning}
Taken together, previous studies have advanced LLM4Code research along three largely disjoint axes: 
(i) designing stronger pre-trained models,
(ii) enhancing reasoning transparency via CoT prompting, and
(iii) developing uncertainty-based or adversarial robustness evaluations.
However, no existing work systematically examines how reasoning structure modulates uncertainty—and how these uncertainty dynamics, in turn, predict robustness outcomes.

Our study fills this gap by conducting a large-scale empirical analysis across diverse datasets and perturbation types, linking CoT reasoning, uncertainty evolution, and robustness in a single interpretive framework.
This integration advances beyond prior descriptive evaluations, offering a principled basis for understanding how LLM4Code models reason and fail, when faced with perturbed inputs.

%% file: Methodolog.tex
\section{Methodology} \label{sec:methodology}

\begin{figure*}[h!tbp]
\centering
\scriptsize
\resizebox{\textwidth}{!}{%
\begin{tikzpicture}[
  node distance=5mm and 6mm,
  font=\scriptsize,
  box/.style={draw, rounded corners, align=left, inner sep=3pt, minimum width=2.85cm, text width=2.85cm},
  smallbox/.style={draw, rounded corners, align=left, inner sep=3pt, minimum width=2.85cm, text width=2.85cm},
  hdr/.style={font=\bfseries},
  arrow/.style={-{Latex[length=1.6mm]}, semithick},
  group/.style={draw, rounded corners, inner sep=4pt, dashed}
]

\node[box] (data) {
  (\textbf{$\mathcal{A}$) Data}\\
  MHPP; BigCodeBench
};

\node[box, right=of data] (prompt) {
  (\textbf{$\mathcal{A}$) Prompts}\\
  No-CoT; CoT\\
  (+ noise-aware tmpl.)
};

\node[box, right=of prompt] (pert) {
  (\textbf{$\mathcal{B}$) Perturbations}\\
  C1--C3; W1--W3; S1\\
  (docstring only)
};

\node[box, below=8mm of prompt] (models) {
  (\textbf{$\mathcal{C}$) Models \& sampling}\\
  DeepSeek / Qwen / CodeLlama\\
  $T\in\{0.5,1.0\}$; $k\in\{1,5,10\}$
};

\node[smallbox, below left=6mm and 4mm of models] (nocot) {
  \textbf{No-CoT}\\
  Code-only
};

\node[smallbox, below right=6mm and 4mm of models] (cot) {
  \textbf{CoT}\\
  Pseudocode $\rightarrow$ Code
};

\node[box, below=7mm of models] (logp) {
  (\textbf{$\mathcal{C}$) Token log-probs}\\
  Entropy; Prob-diff
};

\node[box, below left=7mm and 10mm of logp] (tests) {
  (\textbf{$\mathcal{D}$) Execution eval}\\
  Unit tests\\
  pass/fail
};

\node[box, left=of tests] (metrics) {
  (\textbf{$\mathcal{D}$) Outcome metrics}\\
  Pass@k\\
  RD (robustness)
};

\node[box, below right=7mm and 10mm of logp] (unc) {
  (\textbf{$\mathcal{E}$) Uncertainty}\\
  Early-window agg.\\
  AUROC; Spearman $\rho$
};

\node[box, right=of unc] (traj) {
  (\textbf{$\mathcal{F}$) Trajectory}\\
  Anchors: A1/A2/A3\\
  Deform: L/B/S
};

\draw[arrow] (data) -- (prompt);
\draw[arrow] (prompt) -- (pert);
\draw[arrow] (pert) |- (models);

\draw[arrow] (models) -- (logp);
\draw[arrow] (models) -- (nocot);
\draw[arrow] (models) -- (cot);

\draw[arrow] (logp) -- (tests);
\draw[arrow] (tests) -- (metrics);

\draw[arrow] (logp) -- (unc);
\draw[arrow] (unc) -- (traj);

\node[draw, rounded corners, inner sep=2pt, font=\scriptsize, above=1mm of metrics] (rq12) {RQ1+RQ2};
\draw[dotted] (rq12) -- (metrics);

\node[draw, rounded corners, inner sep=2pt, font=\scriptsize, above=1mm of unc] (rq3) {RQ3};
\draw[dotted] (rq3) -- (unc);

\node[draw, rounded corners, inner sep=2pt, font=\scriptsize, above=1mm of traj] (rq4) {RQ4};
\draw[dotted] (rq4) -- (traj);

\begin{scope}[on background layer]
  \node[group, fit=(data) (prompt) (pert), label={[hdr]above:Inputs}] {};
  \node[group, fit=(models) (nocot) (cot) (logp), label={[hdr]above:Generation}] {};
  \node[group, fit=(metrics) (tests), label={[hdr]above:Evaluation}] {};
  \node[group, fit=(unc) (traj), label={[hdr]above:Analysis}] {};
\end{scope}
\end{tikzpicture}%
}
\vspace{4pt}
\caption{\textbf{Methodology overview and research questions (RQs) mapping.} Steps are labeled $\mathcal{A}$--$\mathcal{F}$. We evaluate CoT and No-CoT prompting on MHPP and BigCodeBench under controlled character (C), word (W), and sentence (S) perturbations. Execution-based metrics (Pass@k, RD) address \textbf{RQ1} (performance impact of CoT) and \textbf{RQ2} (robustness under perturbations). Token-level uncertainty signals (entropy, probability differential) support \textbf{RQ3} (early failure predictiveness via AUROC and Spearman $\rho$). Reasoning–code trajectory analysis around structural anchors addresses \textbf{RQ4}, identifying deformation patterns—lengthening (L), branching (B), and simplification (S).}
\label{fig:method_overview_ieee}
\end{figure*}
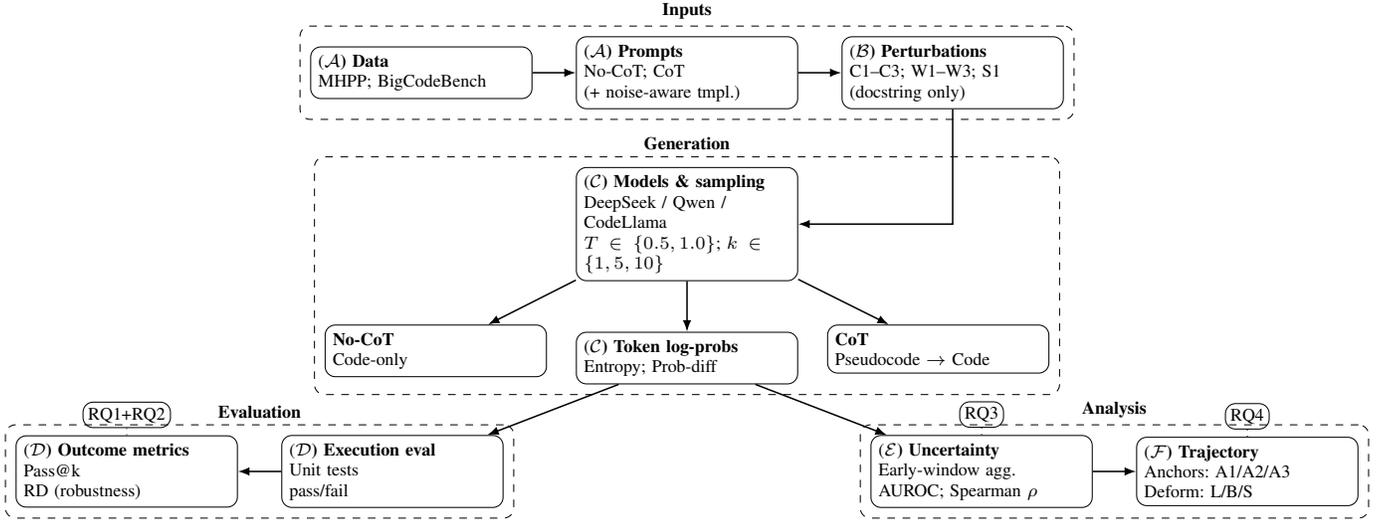

This section presents the methodological framework used to investigate how CoT reasoning influences uncertainty dynamics and robustness in large language models for code generation (LLM4Code). Our design combines large-scale evaluations on diverse benchmarks with fine-grained uncertainty tracking and controlled prompt perturbations. LLMs generate text through an autoregressive decoding process, where tokens are produced sequentially conditioned on previously generated tokens. This sequential property makes it possible to analyze token-level uncertainty over time and to align uncertainty signals with structural events in the generation trajectory.


\subsection{Overview}

Figure~\ref{fig:method_overview_ieee} illustrates our experimental methodology. We systematically evaluate the impact of CoT reasoning on code generation robustness under controlled linguistic perturbations. The methodology comprises three core phases:
\begin{itemize}[leftmargin=*,noitemsep,topsep=0pt]
    \item \textbf{Input preparation}---constructing CoT and No-CoT prompts from two code benchmarks (MHPP and BigCodeBench) and applying character-, word-, and sentence-level perturbations exclusively to documentation strings.
    \item \textbf{Code generation}---executing multiple state-of-the-art LLMs under varying temperature and sampling parameters to produce both reasoning-augmented (CoT) and code-only (No-CoT) solutions.
    \item \textbf{Evaluation and analysis}---assessing functional correctness and robustness, extracting token-level uncertainty signals, and analyzing reasoning trajectory deformations to address our four research questions.
\end{itemize}

This integrated pipeline enables a controlled comparison between CoT and No-CoT prompting, linking correctness and robustness outcomes to uncertainty dynamics and reasoning trajectory stability under distributional noise.

To operationalize our framework, we formulate four research questions (RQs) that examine performance, robustness, uncertainty predictiveness, and trajectory deformation under perturbation. These RQs structure our empirical analysis and connect each component of the pipeline to a concrete evaluative objective.

\subsection{Research Questions (RQs)}
\textbf{\rqone}

\textbf{Rationale.} 
CoT prompting is a popular technique for eliciting explicit reasoning from LLMs, but empirical evidence on its consistent benefits for code generation remains inconclusive. Understanding whether CoT provides universal improvements or exhibits task- and model-specific effects is essential for both theoretical understanding of reasoning-enhanced generation and practical deployment decisions.

\textbf{Objective.} 
Compare CoT vs No-CoT in terms of pass@k across models and datasets without perturbations to determine whether CoT yields uniform gains or model- and task-dependent effects.

\textbf{Approach.} 
For each task, we generate multiple samples under both prompting modes and compute pass@1, pass@5, and pass@10, following common practice in LLM4Code evaluation~\cite{chen2021evaluating}. These values are among the most widely used sampling budgets in prior work, where pass@1 reflects single-attempt reliability, pass@5 captures moderate sampling benefits, and pass@10 approximates the upper bound of practical usage without incurring prohibitive computational cost.


\textbf{\rqtwo}

\textbf{Rationale.} Although CoT is widely used to improve reasoning, prior work has primarily evaluated its effectiveness in clean settings and focused on final accuracy rather than robustness. As a result, we still lack systematic evidence on how CoT compares to No-CoT when inputs are subject to controlled linguistic perturbations. Clarifying whether CoT improves, worsens, or leaves unchanged the robustness of code generation under such perturbations is essential for understanding the reliability of reasoning-based prompting in realistic deployment scenarios.

\textbf{Objective.} Quantify how CoT behaves under character-, word-, and sentence-level perturbations, and whether it amplifies or mitigates robustness degradation.

\textbf{Approach.} We apply seven perturbations (C1–C3, W1–W3, S1), which is shown in Table~\ref{tab:perturbation_definitions}, to all prompts, evaluate both CoT and No-CoT under each perturbation, and measure robustness via Relative Degradation (RD).

\begin{table*}[!ht]
\centering
\caption{Summary of perturbation types used in our robustness evaluation, categorized by linguistic level.}
\label{tab:perturbation_definitions}
\begin{tabular}{@{}llp{14cm}@{}}
\toprule
\textbf{ID} & \textbf{Name} & \textbf{Definition} \\
\midrule
\multicolumn{3}{c}{\textbf{Character-Level Perturbations}} \\
\cmidrule{1-3}
C1 & Case Flip & Randomly flips letter case within selected words (e.g., \texttt{Input → iNpUt}). Simulates inconsistent capitalization. \\
C2 & Adjacent Swap & Swaps two adjacent letters within a word (e.g., \texttt{function → fucntion}). Mimics typographical errors. \\
C3 & Letter Replacement & Replaces characters in multiple words (e.g., \texttt{number → numver}). Disrupts token boundaries. \\
\addlinespace
\midrule
\multicolumn{3}{c}{\textbf{Word-Level Perturbations}} \\
\cmidrule{1-3}
W1 & Synonym Insertion & Inserts synonyms while preserving meaning (e.g., \texttt{sort → sort and arrange}). Increases verbosity. \\
W2 & Synonym Substitution & Replaces words with synonyms (e.g., \texttt{find → locate}). Light paraphrasing altering lexical form. \\
W3 & Inflectional Variation & Applies grammatical variations (e.g., \texttt{returns → returned}). Perturbs surface grammar. \\
\addlinespace
\midrule
\multicolumn{3}{c}{\textbf{Sentence-Level Perturbation}} \\
\cmidrule{1-3}
S1 & Back Translation & Rewrites via translation to another language and back. Alters structure while preserving meaning. \\
\bottomrule
\end{tabular}
\end{table*}

\textbf{\rqthree}

\textbf{Rationale.} While recent studies have explored uncertainty in LLM generation, there is little direct empirical evidence examining whether \emph{early-stage} uncertainty during decoding is actually related to final code correctness, especially in the presence of CoT reasoning. Without such validation, it remains unclear whether early uncertainty can meaningfully inform failure detection or merely reflects transient fluctuations in the generation process.

\textbf{Objective.} Assess whether uncertainty measured in the early part of the generation is informative about eventual pass/fail outcomes.

\textbf{Approach.} Using token-level log probabilities, we compute entropy and probability differential over the decoding steps, derive early-stage aggregates, and evaluate their predictive power.

\textbf{\rqfour}

\textbf{Rationale.} Prior analyses of CoT largely treat reasoning as a black box or focus solely on final outputs, overlooking how intermediate reasoning trajectories evolve and fail. However, understanding where and how perturbations reshape the reasoning--code pathway is crucial for explaining differential robustness between CoT and No-CoT and for designing more stable reasoning-based systems.

\textbf{Objective.} Analyze how CoT reshapes the reasoning–code trajectory, and how perturbations deform this trajectory into patterns such as lengthening, branching, or simplification.

\textbf{Approach.} We align reasoning segments and code segments, compare structural properties (length, stability, handoff position) across CoT and No-CoT, and relate these patterns to robustness and uncertainty measurements.

\subsection{Dataset and Model Selection Process}
To minimize selection bias and enhance reproducibility, we adopted a structured two-stage selection procedure for datasets and models. 
First, we defined high-level selection principles emphasizing task diversity, reasoning complexity, methodological compatibility with token-level uncertainty analysis, and comparability with established LLM4Code benchmarks. 
Second, we operationalized these principles through explicit inclusion and exclusion criteria to systematically identify benchmarks and model families aligned with our research objectives.

\noindent\textbf{Inclusion criteria.}  
For datasets, we required that they (1) target functional code synthesis from natural-language specifications, (2) provide executable unit tests to objectively determine correctness, and (3) contain non-trivial reasoning or realistic programming scenarios such that CoT is potentially beneficial. Based on these criteria, we included MHPP(Mostly Hard Python Problems) and BigCodeBench (BCB): MHPP stresses algorithmic reasoning and multi-step problem solving, while BCB contains tasks that are closer to real-world programming scenarios and involve longer, more context-rich problem descriptions. We did not adopt HumanEval or MBPP as primary benchmarks because these datasets are widely considered relatively simple and may not sufficiently stress the reasoning and robustness aspects that are central to our study. 

For models, we included state-of-the-art open-weight LLM4Code models that (1) are widely recognized or popular in the community, (2) support autoregressive decoding with access to token-level log probabilities, and (3) offer Instruct variants aligned with natural-language prompts. These conditions ensure that we can both evaluate practical performance and perform fine-grained uncertainty analysis. Accordingly, we selected models such as CodeLlama, DeepSeek-Coder, and Qwen.

\noindent\textbf{Exclusion criteria.}  
We excluded datasets that lack executable test suites (which would preclude reliable pass/fail evaluation) or that primarily involve trivial code completion rather than reasoning-based synthesis. We also excluded closed-source models (e.g., GPT-4/5) from RQ3--RQ4 because they do not expose token-level log probabilities required for our uncertainty and trajectory analyses; this is a methodological necessity rather than a performance-based preference.

\noindent\textbf{Rationale for coverage.}  
Overall, MHPP and BigCodeBench cover complementary dimensions of difficulty and realism, while our model set balances representativeness, reproducibility, and analytical tractability for uncertainty-based analysis.

\subsubsection{Benchmark Datasets}
We leverage two challenging benchmark datasets for evaluating the performance, robustness, and uncertainty behavior of LLM4Code models: MHPP (Mostly Hard Python Problems)~\cite{dai2024mhpp} 
and BigCodeBench (BCB)~\cite{zhuo2024bigcodebench}
Both datasets are designed to reflect realistic coding scenarios and provide executable unit tests for automated correctness evaluation, as illustrated in Figure~\ref{fig:method_overview_ieee}.

\noindent\paragraph{\textbf{\emph{MHPP}}}

The MHPP~\cite{dai2024mhpp} dataset contains a curated collection of algorithmically challenging Python programming tasks derived from BigCodeBench. Unlike HumanEval or MBPP, which focus on short and relatively simple problems, MHPP emphasizes multi-step reasoning, non-trivial control flow, and algorithmic decomposition. Each problem includes a natural-language description, a function signature, hidden unit tests, and a reference solution. These properties make MHPP a suitable benchmark for examining whether CoT genuinely improves reasoning-heavy code generation (RQ1) and how perturbations reshape the reasoning trajectory (RQ4). In our experiments, MHPP serves as the primary dataset for analyzing reasoning behavior under both CoT and No-CoT prompting.

\paragraph{\textbf{\emph{BigCodeBench (BCB)}}}
BigCodeBench(BCB)~\cite{zhuo2024bigcodebench} is a comprehensive code-generation benchmark containing long-form natural-language programming tasks with realistic constraints, examples, and edge-case specifications. Compared to MHPP and other earlier small-scale datasets, BCB presents substantially more complex linguistic structure and diverse problem narratives, making it particularly suitable for robustness evaluation under semantic-preserving perturbations. Each task provides a problem description, metadata, a set of unit tests, and one or more ground-truth reference implementations. Since the original BCB subset contains more samples than MHPP (which includes 210 tasks), we randomly select 298 BCB samples to maintain a comparable scale across datasets and avoid imbalance-induced bias in cross-dataset analysis. This controlled sampling ensures that performance differences are not driven by dataset size disparities. This dataset is especially valuable for assessing robustness degradation (RQ2) and studying linguistic variability effects.

\subsubsection{Dataset Structure.}

Each problem instance in MHPP and BigCodeBench (BCB) follows a structured JSON format, but the two datasets differ in their specific field design and metadata organization.

\noindent\textbf{MHPP.}
Each MHPP instance contains the following fields:

\begin{itemize}[leftmargin=*, noitemsep, topsep=2pt]

    \item \textbf{\texttt{task\_id}}: A unique identifier for the task.
    
    \item \textbf{\texttt{function\_name}}: The target function name to implement.
    
    \item \textbf{\texttt{parameters}}: A structured description of the function parameters.
    
    \item \textbf{\texttt{prompt}}: The natural-language problem description, used as model input.
    
    \item \textbf{\texttt{test}}: A set of unit tests used to evaluate correctness via Pass@$k$.

\end{itemize}

\noindent\textbf{BigCodeBench (BCB).}
Each BCB instance contains the following fields:
\begin{itemize}[leftmargin=*, noitemsep, topsep=2pt]

    \item \textbf{\texttt{task\_id}}: A unique identifier for each programming task.
    
    \item \textbf{\texttt{complete\_prompt}}: The full natural-language task description, including instructions and contextual information. This field is directly used as the model input.
    
    \item \textbf{\texttt{entry\_point}}: The required function name that the model must implement.
    
    \item \textbf{\texttt{test}}: A string containing executable unit tests used to evaluate functional correctness.
    
    \item \textbf{\texttt{canonical\_solution}}: A reference implementation that passes all provided tests.
    
    \item \textbf{\texttt{doc\_struct}}: Structured metadata describing the internal organization of the task description.
    
    \item \textbf{\texttt{libs}}: A list of allowed or required libraries for the task.

\end{itemize}

In both datasets, the unit tests serve as the sole correctness criterion for computing Pass@$k$, while the canonical solutions are used only for validation and sanity checking. Although the field names and metadata structures differ, both datasets follow a consistent paradigm: a natural-language specification paired with executable tests and a predefined function interface.

\subsubsection{Model Selection}

To evaluate the impact of CoT on correctness, robustness, and uncertainty, we experiment with four widely used open-source LLM4Code models: DeepSeek-Coder-6.7B, CodeLlama-7B-Instruct, CodeLlama-13B-Instruct, and Qwen2.5-Coder-7B. These models collectively represent three major model families (DeepSeek, LLaMA-based CodeLlama, and Qwen), providing architectural diversity and reducing family-specific bias in our comparisons.



\noindent\textbf{Model Selection Rationale. } Our core analysis focuses on three models (DeepSeek-Coder, Qwen-Coder, CodeLlama) that support both CoT and No-CoT prompting. Additionally, we include GPT-OSS-20B as a \textbf{triangulation model} to verify the generalizability of CoT-related patterns. Since GPT-OSS-20B inherently produces CoT reasoning regardless of prompt configuration, it cannot participate in direct CoT vs. No-CoT comparisons. Instead, it serves as an external reference point, allowing us to assess whether deformation patterns and uncertainty behaviors observed in our core models also manifest in this architecturally distinct, CoT-constrained implementation.

A key criterion in selecting the four core open-source models is the availability of token-level log probabilities, which are required for analyzing early-stage uncertainty in RQ3. For this reason, we intentionally exclude popular closed-source models such as GPT-5, GPT-4, or GPT-4o, whose APIs do not expose per-token probability distributions and therefore do not support uncertainty-trajectory analysis. However, to broaden the scope of our correctness analysis only for RQ1, we further include GPT-5 mini and GPT-5 nano as supplementary models and are not treated as core comparison models. Preliminary inspection also suggests that CodeLlama-7B may struggle with more complex tasks; thus, we include the larger CodeLlama-13B-Instruct to examine whether this gap stems from limited model capacity. Throughout our experiments, we use the Instruct variants rather than base versions, as they are better aligned with natural-language instructions and reasoning-oriented prompts, ensuring that both CoT and No-CoT templates are consistently interpreted across models.

\subsection{Evaluation Metrics}\label{sec:eval-metrics}
We adopt robustness evaluation metrics (i.e., Pass@K and Relative Degradation) widely used in prior work on code generation~\cite{liu2025adversarial, chen2021evaluating,he2024instruction}. Let $k$ denote the number of top-ranked outputs considered, ordered by model confidence or likelihood. A prediction is considered correct if at least one of the top-$k$ outputs passes the reference unit tests. In addition to correctness and robustness, we evaluate the uncertainty and predictive effectiveness of model behaviors using four widely adopted metrics: Entropy~\cite{zhu2025uncertainty}, Probability Differential~\cite{zhu2025uncertainty}, AUROC~\cite{hanley1982meaning, davis2006relationship}, and Spearman~$\rho$~\cite{spearman1961proof, schmid2007multivariate, bedHo2016multivariate}. These metrics quantify how model uncertainty evolves during the generation process and how well early uncertainty correlates with final code correctness.

We report following metrics:
\begin{itemize}
    \item \textbf{Pass@k.} The probability that a correct solution appears among the top-$k$ outputs. Values range between [0, 1] $\subset \mathbb{R}$, with higher scores indicating better performance.
   \item Relative Degradation (RD) metric quantifies robustness loss under perturbations. Given original performance $P_o = \text{Pass@}k_{\text{original}}$ and perturbed performance $P_p = \text{Pass@}k_{\text{perturbed}}$, both in $[0, 1] \subset \mathbb{R}$, RD is computed as:
    \[
    \text{RD} = 
    \begin{cases}
    0 & \text{if } P_o = 0, \\
    \displaystyle\frac{P_o - P_p}{P_o} & \text{otherwise},
    \end{cases}
    \quad \text{with } \text{RD} \in [0, 1] \subset \mathbb{R}.
    \]
    RD value of 0 indicates no degradation and 1 indicates complete failure.
    
     \item \textbf{Entropy ($H_t$).} Uncertainty measured from the model's next-token probability distribution at decoding step $t$:
    \[ 
        H_t = -\sum_{i=1}^{V} p_t(i) \log_2 p_t(i) \quad \text{(bits)},
    \]
    where $V$ is the vocabulary size, and $p_t(i)$ is the softmax probability for token $i$. $H_t$ ranges from 0 (deterministic prediction) to $\log_2 V$ (uniform distribution). In practice, $H_t$ typically falls between 0.2 and 8.0 bits for code generation models, with lower values indicating high confidence and higher values reflecting generation ambiguity.

    \item \textbf{Probability Differential (prob diff).}  
    The gap between the most probable and second most probable tokens:
    \[
        \Delta_t = p_{t}^{(1)} - p_{t}^{(2)},
    \]
    where $p_{t}^{(1)}$ and $p_{t}^{(2)}$ denote the top-$1$ and top-$2$ token probabilities at step $t$. Smaller values indicate higher ambiguity between competing next tokens.

    \item \textbf{Area Under the ROC Curve (AUROC).}  
    AUROC quantifies how well an uncertainty metric separates successful generations from failed ones. Formally, AUROC is the probability that a randomly chosen failure sample has higher uncertainty than a randomly chosen success sample. Values range from 0.5 (complete randomness, i.e., no predictive power) to 1 (perfect discrimination). In practice, we compute AUROC by sweeping a decision threshold over the uncertainty score and constructing the Receiver Operating Characteristic (ROC) curve, where the x-axis is the false positive rate (FPR: the fraction of successful samples incorrectly classified as failures) and the y-axis is the true positive rate (TPR: the fraction of failed samples correctly identified). AUROC corresponds to the area under this curve. By convention, we report AUROC in the range $[0.5,1]$, as values below 0.5 simply indicate an inverted scoring direction, which can be corrected by negating the uncertainty score.


    \item \textbf{Spearman's Rank Correlation ($\rho$).}  
    A non-parametric correlation measure between early-stage uncertainty and final code correctness:
    \[
        \rho
        =
        \frac{
        \sum_{i=1}^{n} \big(R(x_i)-\bar R_x\big)\big(R(y_i)-\bar R_y\big)
        }{
        \sqrt{\sum_{i=1}^{n}\big(R(x_i)-\bar R_x\big)^2}
        \;
        \sqrt{\sum_{i=1}^{n}\big(R(y_i)-\bar R_y\big)^2}
        },
    \]
    where $R(\cdot)$ denotes the rank operator, $x_i$ is the early-stage uncertainty feature for sample $i$, and $y_i\in\{0,1\}$ is the binary correctness label (1 = failure).  
    A positive $\rho$ indicates that higher early uncertainty is monotonically associated with higher failure rates.

\end{itemize}

\subsection{Statistical Analysis and Hypothesis Testing}
\label{sec:hypotheses}
We address our research questions through a unified evaluation pipeline that instruments generation traces with token-level uncertainty signals (entropy and probability differential) across two benchmarks (MHPP and BigCodeBench) and multiple character-, word-, and sentence-level perturbations. By analyzing the temporal evolution of these signals along structurally meaningful points in the reasoning--code trajectory, we examine whether early uncertainty can serve as an indicator of generation failure and how CoT reasoning behaves under distributional perturbations.
To formally evaluate our research questions, we formulate null ($H_0^{x}$) and alternative ($H_1^{x}$) hypotheses. All statistical tests are conducted using a significance level $\alpha = 0.05$. We reject the null hypothesis when $p < \alpha$ and otherwise fail to reject the null hypothesis.\\
\noindent\textbf{(RQ1: Performance Impact of CoT)}
\begin{itemize}
\item \textbf{$H_0^{1}$:} There is no statistically significant difference in Pass@k between CoT and No-CoT prompting.
\item \textbf{$H_1^{1}$:} There is a statistically significant difference in Pass@k between CoT and No-CoT prompting.
\end{itemize}
\noindent\textbf{(RQ2: Robustness Under Perturbations)}
\begin{itemize}
\item \textbf{$H_0^{2}$:} There is no statistically significant difference in Relative Degradation (RD) between CoT and No-CoT under linguistic perturbations.
\item \textbf{$H_1^{2}$:} There is a statistically significant difference in Relative Degradation (RD) between CoT and No-CoT under linguistic perturbations.
\end{itemize}
\noindent\textbf{(RQ3: Early Uncertainty Predictiveness)}
\begin{itemize}
\item \textbf{$H_0^{3}$:} Early-stage uncertainty is not statistically
associated with final code-generation failure.
\item \textbf{$H_1^{3}$:} Early-stage uncertainty is statistically associated with final code-generation failure.
\end{itemize}
\noindent\textbf{(RQ4: Structural Anchor Effects)}
\begin{itemize}
\item \textbf{$H_0^{4a}$:} Linguistic perturbations do not significantly change anchor-aligned spike distances in CoT trajectories.
\item \textbf{$H_1^{4a}$:} Linguistic perturbations significantly change anchor-aligned spike distances in CoT trajectories.
\item \textbf{$H_0^{4b}$:} Deformation pattern frequencies are independent of perturbation type.
\item \textbf{$H_1^{4b}$:} Deformation pattern frequencies depend on
perturbation type.
\end{itemize}
To test these hypotheses, we employ non-parametric statistical methods
appropriate for paired comparisons, association analysis, and
distributional differences. For RQ1 and RQ2, we compare performance
metrics between CoT and No-CoT prompting using the Wilcoxon signed-rank test for paired samples. For RQ3, we evaluate whether early uncertainty signals are associated with generation outcomes using the Spearman rank correlation coefficient and Area Under the Receiver Operating Characteristic Curve (AUROC), which measure monotonic association and predictive discrimination, respectively.
For RQ4, we conduct two complementary analyses capturing both
quantitative shifts in uncertainty localization and categorical
changes in reasoning trajectory structure.
\paragraph{Anchor-aligned spike localization ($H_0^{4a}$).}
We analyze whether perturbations systematically shift the localization
of uncertainty spikes relative to structural anchors. For each
generation trace, we compute the normalized spike distance
\[
\Delta_k = \frac{S - A_k}{T}
\]
where $S$ denotes the position of the first uncertainty spike,
$A_k$ denotes the position of anchor $k$, and $T$ is the total
sequence length. This metric quantifies how early or late instability
occurs relative to structural commitment points in the reasoning
trajectory.
To evaluate $H_0^{4a}$, we compare $\Delta_k$ values between clean and
perturbed prompts using the Wilcoxon signed-rank test for paired
samples. This test is appropriate because the same tasks are evaluated
under both prompt conditions, producing paired observations without
assuming normality of the underlying distributions. Effect sizes are
reported using the Wilcoxon effect size
\[
r = \frac{|Z|}{\sqrt{N}}
\]
where $Z$ is the standardized Wilcoxon statistic and $N$ is the number
of paired observations. As a robustness check, we additionally apply the two-sample Kolmogorov--Smirnov (KS) test to compare the full distributions of spike distances. The KS statistic
\[
D = \sup_x |F_1(x) - F_2(x)|
\]
measures the maximum distance between the empirical cumulative
distribution functions of the two samples and quantifies the magnitude
of distributional differences.
\paragraph{Trajectory deformation patterns ($H_0^{4b}$).}
In addition to spike localization, we analyze structural changes in
reasoning trajectories. Each trajectory is categorized into one of
three deformation types: \emph{lengthening}, \emph{branching}, or
\emph{simplification}. These categories capture how perturbations
alter the structure of the reasoning process.
To evaluate $H_0^{4b}$, we construct contingency tables comparing
deformation frequencies across perturbation families and apply the
Chi-Square Test of Independence ($\chi^2$ test). This test evaluates
whether deformation patterns and perturbation types are statistically
independent. The test is appropriate because both perturbation type
and deformation pattern are categorical variables. Effect sizes are
reported using Cramér's $V$, which measures the strength of association between categorical variables.
Together, these analyses allow us to determine whether perturbations
systematically reshape reasoning trajectories relative to structural
anchors rather than producing random variations in generation behavior.
\paragraph{Effect sizes and statistical decision rules.}
Additionally, we report effect-size measures corresponding to each
statistical test. For $H_0^{1}$, $H_0^{2}$, and $H_0^{4a}$, we report the Wilcoxon effect size $r$. Following conventional thresholds,
$r < 0.1$ indicates a negligible effect (N), $0.1 \le r < 0.3$ a small
effect (S), $0.3 \le r < 0.5$ a medium effect (M), and $r \ge 0.5$ a
large effect (L), reflecting increasing magnitudes of difference
between paired conditions. For $H_0^{3}$, we report the Spearman rank correlation coefficient ($\rho$) and the AUROC to quantify the strength of association and predictive discrimination between early uncertainty signals and generation outcomes. For $H_0^{4a}$, we additionally report the Kolmogorov--Smirnov statistic $D$, which captures distributional differences in spike localization between clean and perturbed prompts. For $H_0^{4b}$, effect sizes are reported using Cramér's $V$, which measures the strength of association between categorical variables in contingency tables.

\subsection{Prompting and Generation Setup}
We employ two prompting conditions—CoT and No-CoT—to isolate the effect of explicit intermediate reasoning on code generation robustness and uncertainty dynamics. This design allows us to compare a minimalist generation strategy (No-CoT), which directly maps specifications to code, against a reasoning-augmented strategy (CoT), which encourages the model to articulate intermediate steps before producing code, a widely studied mechanism in prior work on LLM reasoning.

We apply both prompting modes to all tasks under unperturbed and perturbed evaluation settings, yielding a $2\times2$ design (No-CoT vs. CoT $\times$ clean vs. perturbed inputs). Accordingly, the four prompt templates used in our study correspond to: (1) a base No-CoT template, (2) a base CoT template, (3) a perturbation-aware No-CoT template, and (4) a perturbation-aware CoT template. Importantly, these perturbation-aware templates do not contain corrupted text themselves; rather, they explicitly instruct the model on how to interpret inputs whose docstrings have been perturbed.

\subsubsection{No-CoT prompting}


In the No-CoT setup, we instruct the model to output only the final Python implementation in a single code block, without revealing intermediate reasoning. The prompt enforces strict formatting rules (e.g., exactly one Python block and no explanations). In the perturbation-aware variant, we additionally inform the model that the docstring may contain noise (such as letter-case flips, adjacent swaps, synonym substitutions, or paraphrasing) and explicitly describe these possible perturbation types, while instructing the model to consider the intended semantics of the task. We do not assume that the model can always perfectly identify or ignore such perturbations; rather, these instructions provide additional context to examine how the model responds when made aware that surface-level noise may be present. Concrete examples of these two No-CoT configurations (base and perturbation-aware) are provided in Table~\ref{tab:prompt_examples_no_cot}.

\begin{table*}[t]
\centering
\caption{No-CoT prompt templates used in this study. The templates are shown in abridged form for readability; full prompts are provided in the supplementary materials.}
\label{tab:prompt_examples_no_cot}
\renewcommand{\arraystretch}{1.2}
\begin{tabular}{|p{3.2cm}|p{10.1cm}|}
\hline
\textbf{Condition} & \textbf{Template (abridged)} \\
\hline
\textbf{Base No-CoT} &
\begin{minipage}[t]{10.0cm}
\vspace{0.3em}
\texttt{Role: You are a software programmer.}\\[-0.2em]
\texttt{Task: Complete the function strictly following the signature and description.}\\[-0.2em]
\texttt{Do NOT explain your reasoning.}\\[-0.2em]
\texttt{Output exactly one Python code block and nothing else.}
\vspace{0.3em}
\end{minipage} \\
\hline
\textbf{Perturbation-aware No-CoT} &
\begin{minipage}[t]{10.0cm}
\vspace{0.3em}
\texttt{Role: You are a software programmer.}\\[-0.2em]
\texttt{Task: Complete the function strictly following the signature and description.}\\[-0.2em]
\texttt{Note: The docstring may contain perturbation (case flips, swaps,}\\[-0.2em]
\texttt{synonym substitutions, or paraphrasing). Infer the intended}\\[-0.2em]
\texttt{behavior regardless of such surface changes.}\\[-0.2em]
\texttt{Do NOT explain your reasoning; output exactly one Python block.}
\vspace{0.3em}
\end{minipage} \\
\hline
\end{tabular}
\end{table*}

\subsubsection{CoT prompting}


Under the CoT condition, the model must first produce a \textit{Pseudocode} section describing the reasoning steps before generating the final code block. We impose this two-stage structure to explicitly separate high-level reasoning from low-level implementation, which allows us to examine (i) how perturbations reshape the intermediate reasoning trajectory and (ii) whether instability in these early reasoning steps anticipates downstream code failures. The CoT template, defined in Table~\ref{tab:prompt_examples_cot}, explicitly enforces this two-stage output format and prohibits any extraneous commentary. 

Similar to the No-CoT case, the perturbation-aware CoT template does not alter the task itself; instead, it explicitly describes possible forms of docstring perturbations (e.g., case flips, swaps, synonym substitutions, or paraphrasing) and instructs the model to consider the intended semantics of the task. We do not assume that the model can always perfectly identify or ignore such perturbations; rather, these instructions serve to test how the model responds when made aware that surface-level noise may be present.

\begin{table*}[t]
\centering
\caption{CoT prompt templates used in this study. The templates are shown in abridged form for readability; full prompts are provided in the supplementary materials.}
\label{tab:prompt_examples_cot}
\renewcommand{\arraystretch}{1.2}
\begin{tabular}{|p{3.2cm}|p{10.1cm}|}
\hline
\textbf{Condition} & \textbf{Template (abridged)} \\
\hline
\textbf{Base CoT} &
\begin{minipage}[t]{10.0cm}
\vspace{0.3em}
\texttt{Role: You are a software programmer.}\\[-0.2em]
\texttt{Task: Complete the function strictly following the signature and description.}\\[-0.2em]
\texttt{Use a Chain-of-Thought approach.}\\[-0.2em]
\texttt{Write a ``Pseudocode:'' section (no code fence),}\\[-0.2em]
\texttt{then output exactly one Python code block.}\\[-0.2em]
\texttt{Stop after the closing ``` .}
\vspace{0.3em}
\end{minipage} \\
\hline
\textbf{Perturbation-aware CoT} &
\begin{minipage}[t]{10.0cm}
\vspace{0.3em}
\texttt{Role: You are a software programmer.}\\[-0.2em]
\texttt{Task: Complete the function strictly following the signature and description.}\\[-0.2em]
\texttt{The docstring may contain noise (case flips, swaps,}\\[-0.2em]
\texttt{synonym substitutions, or paraphrasing). Always infer}\\[-0.2em]
\texttt{the intended behavior.}\\[-0.2em]
\texttt{Use Chain-of-Thought: write ``Pseudocode:'' first,}\\[-0.2em]
\texttt{then exactly one Python code block.}
\vspace{0.3em}
\end{minipage} \\
\hline
\end{tabular}
\end{table*}

\subsubsection{Perturbation handling}


The perturbations themselves are not embedded into the prompt templates. Instead, we apply perturbations directly to the task description (the docstring) of each dataset instance, while keeping the function signature unchanged. We preserve the signature to maintain test executability and to isolate the effect of linguistic perturbations on problem understanding—rather than on interface-level changes—so that any performance differences can be attributed to how models interpret perturbed specifications under otherwise identical coding requirements. The perturbation-aware templates do not contain corrupted text; rather, they provide consistent guidance to the model on how to handle potentially corrupted inputs, ensuring comparability across CoT and No-CoT settings.
\textbf{Table~\ref{tab:perturbation_definitions}} summarizes the perturbation operations used in this study, including their linguistic level, operational definition, and an illustrative example.

\subsubsection{Generation configuration}

For each (model, prompting mode, perturbation type), we sample multiple completions at temperatures 0.5 and 1.0. All generated programs are executed against the official unit tests to determine correctness. Additionally, token-level log probabilities are captured at every decoding step to support the uncertainty-trajectory analyses conducted in RQ3 and RQ4.

\subsection{Anchor Definitions and Uncertainty Alignment}
\label{sec:anchors}

To analyze how uncertainty evolves along the generation trajectory, we introduce a set of \emph{structural anchors} that mark semantically meaningful transitions during code generation. These anchors serve as \emph{temporal reference points} rather than causal triggers, enabling alignment of uncertainty signals across heterogeneous generation traces.



Let a generation trace be represented as an \textbf{ordered token sequence}
\[
\mathbf{x} = (x_1, x_2, \dots, x_T) \in \mathcal{V}^T,
\]
where $\mathcal{V}$ is the token vocabulary and $T$ is the trace length. 
The sequence representation preserves the \textbf{temporal ordering} essential 
for anchor definitions and uncertainty alignment, because:
\begin{itemize}[leftmargin=*,noitemsep,topsep=2pt]
    \item Generation is autoregressive: $x_t$ depends on $x_1, \dots, x_{t-1}$; 
    \item Anchors ($A_1, A_2, \text{and\;} A_3$) rely on positional predicates that identify ``first occurrences'' of specific structural patterns;
    \item Uncertainty signals $u_t$ are aligned to specific positions $t$.
\end{itemize}
At each position $t$, we observe a token-level uncertainty signal $u_t$ (e.g., entropy derived from output probabilities). An anchor is defined as a token position $A_k \in \{1,\dots,T\}$ associated with a specific structural event. To enable comparison across traces of different lengths, we compute anchors over \textbf{325,120} generation traces collected from two models (DeepSeek and Qwen) on MHPP ($210$ tasks) and BigCodeBench ($298$ tasks), under eight input conditions, two prompting modes (CoT and No-CoT), and two temperature settings. In section~\ref{sec:anchor}, we show the effectiveness of these three anchors.
Since the RQ4 analysis includes all sampled completions under the pass@10 evaluation protocol, each problem-setting combination contributes ten generation traces, yielding a total of $(210+298)\times 8\times 2\times 2\times 2\times 10 = 325,120$ traces. Here, each trace corresponds to one sampled model completion. We therefore use the normalized anchor position $\hat{A}_k = A_k / T$ to support meaningful comparison across variable-length outputs.
The anchors correspond to recurring structural commitment points in the reasoning-to-code trajectory. In particular, $A_1$ marks the reasoning-to-code transition, while $A_2$ and $A_3$ capture later stages of symbolic commitment and algorithmic articulation. These anchors are introduced as operational structural reference points rather than assumed causal milestones. Because $A_2$ and $A_3$ may appear less self-evident than $A_1$, we explicitly evaluate their effectiveness in Section~\ref{sec:anchor}, where we show that they consistently align with structurally meaningful positions along the generation trajectory.

\subsubsection{\textbf{$A_1$}: Reasoning--Code Transition Anchor.}
The first anchor, denoted as $A_1$, marks the transition from natural-language reasoning to executable code generation. Operationally, $A_1$ is defined as the position of the first token that indicates the start of a code block (e.g., a Markdown code fence). Formally,
\[
A_1 = \min \{\, t \mid \mathbb{I}_{\text{code}}(x_t) = 1 \,\},
\]
where $\mathbb{I}_{\text{code}}(x_t)$ is an indicator function that equals $1$ if token $x_t$ corresponds to a code block start and $0$ otherwise. $A_1$ provides a clear temporal boundary separating reasoning-oriented generation from code-level execution. This transition is expected to introduce distributional shifts in token prediction, as the model moves from natural-language reasoning to structured program synthesis, making it a natural reference point for examining uncertainty changes.\\

\subsubsection{\textbf{$A_2$:} Symbolic Commitment Anchor.}

The second anchor, $A_2$, captures the point at which the model makes an explicit symbolic commitment, such as introducing a program identifier (e.g., a variable or data structure) in the reasoning segment that is subsequently reused in the generated code. We view them as markers of early symbolic grounding: once an identifier introduced in reasoning reappears in code, subsequent generation may become increasingly tied to that previously established symbol. This makes $A_2$ a useful reference point for examining whether uncertainty spikes tend to emerge around moments where symbolic grounding becomes salient.

Operationally, $A_2$ is defined as the \textbf{earliest occurrence in the reasoning segment} of any identifier that is reused at least $\lambda$ times in the subsequent code segment. Here, $\lambda$ denotes a minimum reuse threshold used to distinguish incidental mentions from identifiers that play a more structural role in the generated program. Identifiers that appear only once in code may correspond to local or transient expressions, whereas identifiers that are reused multiple times are more likely to participate in variable tracking, data access, or control flow organization. We therefore use repeated occurrence as an observable signal of symbolic commitment.

Let $\mathcal{S}$ denote the set of identifiers that appear at least $\lambda$ times after the code transition point $A_1$:
\[
\mathcal{S} = \left\{ s \;\middle|\; \sum_{t > A_1} \mathbb{I}_s(x_t) \ge \lambda \right\},
\]
where $\mathbb{I}_s(x_t)$ indicates whether token $x_t$ corresponds to identifier $s$. The anchor $A_2$ is then defined as the earliest occurrence of any identifier $s \in \mathcal{S}$ in the reasoning segment:
\[
A_2 = \min \left\{\, t < A_1 \;\middle|\; \exists s \in \mathcal{S},\; \mathbb{I}_s(x_t)=1 \,\right\}.
\]

The choice of $\lambda$ controls the strictness of this definition. If $\lambda=1$, the criterion becomes too permissive, since any identifier that later appears once in code would qualify, making $A_2$ less effective at distinguishing structural commitments from incidental mentions. By contrast, larger thresholds (e.g., $\lambda \ge 3$) make the anchor increasingly restrictive and bias it toward longer or more repetitive solutions, reducing anchor coverage in shorter or simpler programs. In our experiments, we set $\lambda=2$, which corresponds to the minimal non-trivial criterion for identifier reuse. This choice is also consistent with the scale of our benchmark tasks, where identifier reuse is typically limited rather than highly repetitive, so a stricter threshold would unnecessarily exclude many valid anchor instances. For more complex code-generation settings involving longer reasoning traces or denser symbolic reuse, a larger $\lambda$ may be appropriate. We therefore treat $\lambda=2$ as a task-sensitive but practical default that preserves symbolic reuse while maintaining sufficient anchor coverage for cross-trace analysis.

Notably, $A_2$ is a \emph{conditional anchor}: it may be absent when no identifier introduced in reasoning is later reused sufficiently often in the code. This anchor therefore provides a reference point for evaluating whether uncertainty spikes tend to occur near moments where symbolic commitments formed during reasoning propagate into executable code.

\subsubsection{\textbf{$A_3$:} Algorithmic Structure Anchor.}
The third anchor, $A_3$, marks the onset of algorithmic structure within the generated code. Specifically, $A_3$ corresponds to the first appearance of a control-flow or structural construct (e.g., function definitions, loops, or conditionals). Let $\mathcal{C}$ denote a set of such control keywords $\{\texttt{def}, \texttt{for}, \texttt{while}, \texttt{if}, \texttt{else}, \texttt{return}, \dots\}$; then
\[
A_3 = \min \{\, t > A_1 \mid \mathbb{I}_{\mathcal{C}}(x_t)=1 \,\},
\]
where $\mathbb{I}_{\mathcal{C}}(x_t)$ indicates whether token $x_t$ belongs to $\mathcal{C}$. Since $t > A_1$, this anchor specifically captures algorithmic structure formation within the code segment, marking the transition from surface-level syntax generation to structured logic. The introduction of control-flow constructs typically requires the model to instantiate a coherent algorithmic plan, which may increase decision complexity and therefore affect uncertainty dynamics around this point.\\



\subsubsection{Anchor-Aligned Uncertainty Measure.}
To quantify temporal alignment between uncertainty spikes and structural transitions, we define an anchor-aligned normalized distance. Let $S \in \{1,\dots,T\}$ denote the position of the first detected uncertainty spike (e.g., where $u_t$ exceeds threshold $\tau$). For each anchor $A_k$ ($k \in \{1,2,3\}$), we compute
\[
\Delta_k = \frac{S - A_k}{T} \in [-1, 1],
\]
where normalization by $T$ enables comparison across traces of different lengths. $\Delta_k \approx 0$ indicates the uncertainty spike occurs near anchor $A_k$; $\Delta_k < 0$ means the spike precedes $A_k$, and $\Delta_k > 0$ means it follows $A_k$. Throughout our analysis, anchors serve as temporal reference points rather than causal mechanisms; alignment patterns are interpreted descriptively, not causally. 

Empirical distributions of $\Delta_k$ are reported in Figure~\ref{fig:spike_anchor_alignment} in section~\ref{sec:anchor}, where we analyze how uncertainty aligns with structural anchors under different prompting conditions.



%% file: Results.tex
\section{Results} \label{sec:results}
In this section, we present the empirical results addressing our four research questions (RQ1–RQ4), which examine (i) whether CoT improves code-generation performance, (ii) whether CoT enhances robustness under perturbed prompts, (iii) whether early-stage uncertainty can predict downstream code-generation failure, and (iv) how CoT reasoning trajectories are reshaped under perturbations. Our analyses build directly on the methodology, datasets, perturbation design, and evaluation setup described in Section~\ref{sec:methodology}.
To ensure a theory-driven evaluation, each RQ is examined through formal hypothesis testing as defined in Section~\ref{sec:hypotheses}. Specifically, RQ1 evaluates $H_0^{1}$ by comparing Pass@k between CoT and No-CoT under clean settings using non-parametric paired tests. RQ2 evaluates $H_0^{2}$ by analyzing Relative Degradation (RD) under controlled perturbations and testing whether robustness differs between prompting modes. RQ3 evaluates $H_0^{3}$ by assessing the association between early-stage uncertainty and final correctness using Spearman correlation and AUROC. 
RQ4 first evaluates the effectiveness of the proposed structural anchors by testing whether perturbations significantly shift anchor-aligned spike localization ($H_0^{4a}$). 
After establishing the validity of anchors as structural reference points, we examine whether perturbations reshape reasoning trajectories by inducing deformation patterns that are statistically associated with generation outcomes ($H_0^{4b}$).
For each RQ, we report effect sizes and statistical significance where applicable, and conclude by explicitly stating whether we reject or fail to reject the corresponding null hypothesis.


We conduct a comprehensive evaluation using the MHPP and BigCodeBench (BCB) benchmarks and their perturbed variants, measuring both nominal performance (Pass@k) and robustness (Relative Degradation, RD), and applying Wilcoxon signed-rank tests to assess statistical significance.
For uncertainty analysis, we compute token-level entropy and probability-differential curves and evaluate predictive power via AUROC and rank correlations. Finally, we qualitatively and quantitatively examine CoT trajectory dynamics through uncertainty localization and structural patterns of chain deformation.


\subsection{\emph{\rqone}}


We evaluate $H_0^{1}$ by comparing Pass@k between CoT and No-CoT under clean prompts across models, datasets, sampling budgets, and temperatures. Across all evaluated models and datasets, CoT does not lead to consistent performance improvements. Instead, its effect varies substantially by model family, task type, and sampling temperature.

\subsubsection{CodeLlama: CoT consistently degrades performance}
As shown in Table~\ref{tab:pass@k_unperturbed_test_mhpp_CodeLlama-13b-Instruct-hf}, CoT consistently reduces the performance of CodeLlama-13B on the MHPP dataset.
At T = 0.5, pass@1 drops sharply from 17.1\% to 8.1\%, and even at pass@10 the CoT setting remains worse (31.4\% → 25.2\%).
Similar patterns are observed on the BCB dataset, as well as in our evaluation of the CodeLlama-7B-Instruct-hf model on both benchmarks.

These consistent declines suggest that CoT has a uniformly negative impact on CodeLlama. A likely explanation is that CodeLlama lacks explicit reasoning supervision, causing it to treat CoT chains as noisy context rather than as a meaningful planning structure~\cite{roziere2023code, shi2023large}. The longer reasoning prefix increases variance without yielding better task decomposition, ultimately leading to degraded accuracy. Based on this finding, we won't discuss CodeLlama in following research questions.

\begin{table}[!ht]
\centering
\caption{Performance of CodeLlama-13b-Instruct-hf on MHPP dataset (unperturbed). Results show Pass@$k$ rates for different prompting strategies and temperatures.}
\label{tab:pass@k_unperturbed_test_mhpp_CodeLlama-13b-Instruct-hf}
\begin{tabular}{@{}lccc@{}}
\toprule
\multicolumn{1}{c}{\textbf{Configuration}} & \textbf{Pass@1} & \textbf{Pass@5} & \textbf{Pass@10} \\
\midrule
\multicolumn{4}{c}{\textbf{Temperature $T=0.5$}} \\
\cmidrule{1-4}
No-CoT & 17.1 & 25.2 & 31.4 \\
CoT    & 8.1 & 21.9 & 25.2 \\
\addlinespace
\midrule
\multicolumn{4}{c}{\textbf{Temperature $T=1.0$}} \\
\cmidrule{1-4}
No-CoT & 10.5 & 22.4 & 29.5 \\
CoT    & 10.0 & 20.0 & 27.1 \\
\bottomrule
\end{tabular}
\end{table}

\subsubsection{DeepSeek: CoT exhibits dataset-dependent behavior}
DeepSeek shows a clear dependence on task type, with CoT providing little benefit on MHPP but yielding consistent improvements on BCB. On the MHPP dataset (Table~\ref{tab:pass@k_unperturbed_test_mhpp_deepseek-coder-6.7b-instruct}), CoT does not improve accuracy: at T = 0.5, pass@1 decreases from 25.2\% to 21.4\%, and at T = 1.0 it drops from 21.4\% to 20.5\%, while pass@5 and pass@10 remain nearly unchanged. These results suggest that CoT offers limited value for DeepSeek on deterministic algorithmic tasks.


\begin{table}[!ht]
\centering
\caption{Performance of deepseek-coder-6.7b-instruct on MHPP dataset (unperturbed). Results show Pass@$k$ rates for different prompting strategies and temperatures.}
\label{tab:pass@k_unperturbed_test_mhpp_deepseek-coder-6.7b-instruct}
\begin{tabular}{@{}lccc@{}}
\toprule
\multicolumn{1}{c}{\textbf{Configuration}} & \textbf{Pass@1} & \textbf{Pass@5} & \textbf{Pass@10} \\
\midrule
\multicolumn{4}{c}{\textbf{Temperature $T=0.5$}} \\
\cmidrule{1-4}
No-CoT & 25.2 & 33.3 & 35.7 \\
CoT    & 21.4 & 33.3 & 35.7 \\
\addlinespace
\multicolumn{4}{c}{\textbf{Temperature $T=1.0$}} \\
\cmidrule{1-4}
No-CoT & 21.4 & 32.9 & 38.1 \\
CoT    & 20.5 & 32.4 & 38.6 \\
\bottomrule
\end{tabular}
\end{table}

In contrast, the pattern reverses on the BCB dataset. As shown in Table~\ref{tab:pass@k_unperturbed_test_bcb_deepseek-coder-6.7b-instruct}, CoT consistently improves performance: at T = 0.5, pass@1 increases from 7.4\% to 10.4\%, and pass@10 improves from 11.1\% to 17.1\%; similar gains appear at T = 1.0. This suggests that CoT is beneficial when DeepSeek must interpret multi-step natural-language instructions, where chain-based decomposition can support more coherent task planning.

\begin{table}[!ht]
\centering
\caption{Performance of deepseek-coder-6.7b-instruct on BigCodeBench dataset (unperturbed). Results show Pass@$k$ rates for different prompting strategies and temperatures.}
\label{tab:pass@k_unperturbed_test_bcb_deepseek-coder-6.7b-instruct}
\begin{tabular}{@{}lccc@{}}
\toprule
\multicolumn{1}{c}{\textbf{Configuration}} & \textbf{Pass@1} & \textbf{Pass@5} & \textbf{Pass@10} \\
\midrule
\multicolumn{4}{c}{\textbf{Temperature $T=0.5$}} \\
\cmidrule{1-4}
No-CoT & 7.4 & 10.7 & 11.1 \\
CoT    & 10.4 & 15.8 & 17.1 \\
\addlinespace
\multicolumn{4}{c}{\textbf{Temperature $T=1.0$}} \\
\cmidrule{1-4}
No-CoT & 7.4 & 12.8 & 14.1 \\
CoT    & 10.1 & 16.1 & 17.8 \\
\bottomrule
\end{tabular}
\end{table}

\subsubsection{Qwen: CoT benefits vary with instruction explicitness}

Qwen shows a clear but context-dependent benefit from CoT, and this variation is closely related to how explicitly the task is specified in the prompt. In our setup, the prompts themselves are not perturbed: both conditions use the original task descriptions from the dataset. The difference between the two prompt variants lies in whether the instruction includes \emph{perturbation definitions}, i.e., short explanatory descriptions that clarify the intended perturbation operation. Although these definitions are introduced as part of the prompt template design, their main effect here is to increase instruction explicitness rather than to test robustness under perturbed inputs.

Under the standard prompt setting, CoT consistently improves performance on MHPP (Table~\ref{tab:pass@k_unperturbed_test_mhpp_Qwen2.5-Coder-7B-Instruct}). At $T=0.5$, CoT improves pass@1 from 26.2\% to 27.6\% and pass@10 from 36.2\% to 43.8\%; at $T=1.0$, the gains are larger, with pass@1 increasing from 21.0\% to 25.2\% and pass@10 from 35.7\% to 44.3\%. A similar overall pattern is also observed on BCB. These results suggest that when task instructions remain relatively concise, Qwen can use CoT to better organize semantic interpretation and intermediate planning, thereby improving final code generation.

\begin{table}[!ht]
\centering
\caption{Performance of Qwen2.5-Coder-7B-Instruct on MHPP dataset (unperturbed). Results show Pass@$k$ rates for different prompting strategies and temperatures.}
\label{tab:pass@k_unperturbed_test_mhpp_Qwen2.5-Coder-7B-Instruct}
\begin{tabular}{@{}lccc@{}}
\toprule
\multicolumn{1}{c}{\textbf{Configuration}} & \textbf{Pass@1} & \textbf{Pass@5} & \textbf{Pass@10} \\
\midrule
\multicolumn{4}{c}{\textbf{Temperature $T=0.5$}} \\
\cmidrule{1-4}
No-CoT & 26.2 & 33.3 & 36.2 \\
CoT    & 27.6 & 39.0 & 43.8 \\
\addlinespace
\multicolumn{4}{c}{\textbf{Temperature $T=1.0$}} \\
\cmidrule{1-4}
No-CoT & 21.0 & 30.0 & 35.7 \\
CoT    & 25.2 & 37.1 & 44.3 \\
\bottomrule
\end{tabular}
\end{table}

However, this advantage becomes much smaller when the prompt is augmented with perturbation definitions, which make the instruction itself more explicit (Table~\ref{tab:pass@k_unperturbed_test_mhpp_Qwen2.5-Coder-7B-Instruct_with_perturbation_definitions}). Under this more explicit instruction setting, the no-CoT baseline becomes substantially stronger, and the additional gains from CoT are reduced. At $T=0.5$, CoT slightly decreases pass@1 (32.4\% $\rightarrow$ 31.0\%) while only marginally improving pass@10 (41.4\% $\rightarrow$ 43.3\%). At $T=1.0$, CoT improves pass@1 (27.1\% $\rightarrow$ 29.0\%) but slightly underperforms at pass@10 (45.7\% $\rightarrow$ 44.8\%). This pattern suggests that once perturbation definitions already provide stronger task guidance, CoT contributes less additional structure. In such cases, the longer reasoning trajectory may offer only limited benefit while still introducing more opportunities for drift or error.

\begin{table}[!ht]
\centering
\caption{Performance of Qwen2.5-Coder-7B-Instruct on MHPP dataset with perturbation-aware prompting. Results show Pass@$k$ rates when perturbation definitions are included in the prompt template.}
\label{tab:pass@k_unperturbed_test_mhpp_Qwen2.5-Coder-7B-Instruct_with_perturbation_definitions}
\begin{tabular}{@{}lccc@{}}
\toprule
\multicolumn{1}{c}{\textbf{Configuration}} & \textbf{Pass@1} & \textbf{Pass@5} & \textbf{Pass@10} \\
\midrule
\multicolumn{4}{c}{\textbf{Temperature $T=0.5$}} \\
\cmidrule{1-4}
No-CoT & 32.4 & 38.1 & 41.4 \\
CoT    & 31.0 & 40.0 & 43.3 \\
\addlinespace
\multicolumn{4}{c}{\textbf{Temperature $T=1.0$}} \\
\cmidrule{1-4}
No-CoT & 27.1 & 40.5 & 45.7 \\
CoT    & 29.0 & 41.4 & 44.8 \\
\bottomrule
\end{tabular}
\end{table}

Overall, these results suggest that CoT is most useful when the prompt leaves more room for interpretation, but its incremental value diminishes when perturbation definitions make the instruction more explicit. In other words, part of the observed benefit of CoT may come from compensating for underspecified task instructions rather than from reasoning alone.

Finally, we observe a consistent interaction between sampling temperature and the sampling budget $k$. In our LLM4Code setting, the temperature parameter rescales the model’s next-token distribution before sampling. Lower temperature ($t=0.5$) sharpens the distribution, making high-probability tokens more dominant and reducing stochastic exploration. Higher temperature ($t=1.0$) flattens the distribution, increasing the likelihood of sampling lower-ranked tokens and thereby producing more diverse reasoning and code trajectories. At small $k$ (e.g., pass@1), lower temperature generally yields higher accuracy because it reduces stochastic variation in the early generation steps and stabilizes token choices that strongly affect the final outcome. With only one attempt, such determinism is often advantageous. As $k$ increases, however, the advantage gradually shifts: higher temperature often matches or exceeds lower temperature at pass@5 and pass@10. Although individual trajectories become noisier, the greater diversity across samples increases the probability that at least one candidate follows a correct reasoning path. Thus, diversity becomes beneficial when multiple attempts are available, allowing sampling-based exploration to compensate for increased variability.

\subsubsection{GPT-5 Series: CoT Degrades Performance under Explicit Instruction Following}

To complement the evaluation on open-source LLM4Code models, we additionally examine two proprietary GPT-style models (GPT-5-nano and GPT-5-mini) under the same CoT and No-CoT prompting conditions. These models are included exclusively for RQ1, as their APIs do not expose token-level probability distributions and therefore do not support the uncertainty and trajectory analyses required in later research questions.

Across both datasets and model sizes, the GPT-5 series consistently performs worse when CoT prompting is enabled. 
Table~\ref{tab:gpt5_rq1} summarizes the pass@k results. On the MHPP benchmark, GPT-5-nano exhibits a clear performance degradation under CoT: pass@1 drops from 0.219 to 0.176, pass@5 from 0.433 to 0.348, and pass@10 from 0.471 to 0.395. A similar trend is observed on BigCodeBench, where No-CoT outperforms CoT across all sampling budgets (pass@1: 0.081 $\rightarrow$ 0.067; pass@5: 0.191 $\rightarrow$ 0.158; pass@10: 0.221 $\rightarrow$ 0.205). GPT-5-mini shows the same behavior on MHPP, with pass@1 decreasing from 0.552 under No-CoT to 0.490 under CoT.

These results suggest that, for GPT-style models, explicit reasoning does not function as an effective planning aid for code generation. A plausible explanation lies in the training and alignment regime of GPT-5 models. Through extensive supervised fine-tuning and reinforcement learning, these models are strongly optimized for instruction following and direct code synthesis. As a result, they already internalize latent planning mechanisms that allow them to map well-specified task descriptions directly to executable code. Forcing an explicit CoT stage in this context duplicates internal planning, lengthens the generation trajectory, and introduces additional points of commitment that do not improve—and may even interfere with—final correctness.

From the perspective of our anchor-based framework, GPT-5 models under No-CoT prompting appear to bypass an explicit reasoning--code transition altogether, committing early and directly to code generation. Introducing CoT reintroduces a fragile reasoning--code transition (A1) that is otherwise unnecessary, increasing exposure to trajectory deformation through lengthening or premature branching without providing additional semantic clarification.

Importantly, this behavior does not contradict earlier findings in this section; rather, it reinforces our central claim that the effectiveness of CoT is contingent on model architecture and training characteristics. While some models benefit from CoT when task specifications are underspecified or linguistically complex, instruction-aligned GPT-style models demonstrate that CoT can be actively harmful when direct generation is already well-calibrated.

Given that GPT-5 models (i) consistently underperform under CoT in RQ1, and (ii) do not expose token-level signals required for uncertainty and trajectory analyses, we exclude them from subsequent investigations (RQ2--RQ4). The remainder of this paper therefore focuses on open-source LLM4Code models, where both correctness outcomes and internal generation dynamics can be jointly examined.

\subsection{Statistical Comparison between CoT and No-CoT}
To examine whether the performance difference between CoT and No-CoT prompting is statistically significant, we conduct a Wilcoxon signed-rank test on paired results. The Wilcoxon signed-rank test yields $W=698.5$ with $p=0.705$. The corresponding effect size is $r=0.160$. Since the p-value exceeds the conventional significance threshold ($p<0.05$), the result indicates that the difference between CoT and No-CoT prompting is not statistically significant. Moreover, the small effect size suggests that the magnitude of the performance difference between the two prompting strategies is limited. Overall, these results indicate that although numerical differences may appear between CoT and No-CoT prompting, the observed gap remains modest and statistically indistinguishable under paired comparison.

\begin{findingbox}{Hypothesis Decision (RQ1)}
We reject $H_0^{1}$. The heterogeneous performance effects observed across models and datasets indicate that CoT and No-CoT do not exhibit equivalent behavior, and that CoT’s impact is contingent rather than uniformly beneficial.
\end{findingbox}

\begin{table}[t]
\centering
\caption{Performance of GPT-5 models under CoT and No-CoT prompting (RQ1). Results show Pass@$k$ (P@$k$) rates on MHPP and BigCodeBench.}
\label{tab:gpt5_rq1}
\begin{tabular}{@{}lllcc@{}}
\toprule
\multicolumn{1}{c}{\textbf{Model}} & \textbf{Dataset} & \textbf{Configuration} & \textbf{P@1} & \textbf{P@5 / P@10} \\
\midrule
\multicolumn{5}{c}{\textbf{GPT-5-nano}} \\
\cmidrule{1-5}
& MHPP & No-CoT & 21.9 & 43.3 / 47.1 \\
&      & CoT    & 17.6 & 34.8 / 39.5 \\
\cmidrule{2-5}
& BCB  & No-CoT & 8.1 & 19.1 / 22.1 \\
&      & CoT    & 6.7 & 15.8 / 20.5 \\
\addlinespace
\midrule
\multicolumn{5}{c}{\textbf{GPT-5-mini}} \\
\cmidrule{1-5}
& MHPP & No-CoT & 55.2 & -- \\
&      & CoT    & 49.0 & -- \\
\bottomrule
\end{tabular}
\end{table}

\subsection{\emph{\rqtwo}}


We evaluate $H_0^{2}$ by comparing Relative Degradation (RD) between CoT and No-CoT across perturbation families to determine whether prompting mode influences robustness.

Our original goal in RQ2 was to determine whether CoT provides better robustness than No-CoT under perturbed prompts. However, the empirical evidence does not support a consistent robustness advantage for either mode. Across models, datasets, and temperatures, we do not observe a clear pattern in which CoT invariably yields smaller or larger degradation than No-CoT. 

The statistical results confirm that robustness differences between CoT and No-CoT vary substantially across models and datasets. For DeepSeek-Coder, significant differences are observed on both MHPP and BigCodeBench ($p<0.001$), with moderate-to-large effect sizes, indicating that the two prompting modes respond differently to perturbations. In contrast, for Qwen-Coder on MHPP the difference is not statistically significant ($p=0.442$), suggesting that the two modes exhibit comparable robustness behavior under this dataset.

Instead, the data reveal that CoT and No-CoT exhibit different sensitivity profiles across perturbation families, meaning that robustness differences depend more on which perturbation is applied than on whether CoT is enabled.

To characterize these differences, we aggregate absolute relative degradation ($|RD|$) by perturbation type for each generation mode. Table~\ref{tab:rd_severity_comparison} shows the ranking for both No-CoT and CoT prompting. Under No-CoT, word-level synonym substitution (W2) produces the largest average degradation, followed by stronger character-level perturbations (C2, C3). Character-level edits alter token boundaries or distort key words used to infer the task, while W2 changes the lexical items the model relies on when directly mapping the prompt to code.

In contrast, under CoT the largest degradation occurs for inflectional variation (W3) and multi-word character replacement (C3). Unlike No-CoT, CoT introduces a natural-language reasoning chain before producing code. A reasonable interpretation is that morphological changes and strong character-level corruption disrupt the grammatical and structural cues used to maintain a coherent reasoning sequence. Meanwhile, perturbations such as back-translation (S1) preserve high-level meaning and therefore tend to produce smaller degradation under CoT.
\begin{table}[!ht]
\centering
\caption{Mean absolute Relative Degradation ($|RD|$) by perturbation type. Values are aggregated across models and datasets, with rankings shown for each prompting mode.}
\label{tab:rd_severity_comparison}
\begin{tabular}{@{}ccccc@{}}
\toprule
\multicolumn{1}{c}{\textbf{Perturbation}} & \textbf{No-CoT $|RD|$} & \textbf{Rank} & \textbf{CoT $|RD|$} & \textbf{Rank} \\
\midrule
C1 & 9.71 & 7 & 12.67 & 4 \\
C2 & 15.81 & 2 & 16.70 & 3 \\
C3 & 14.46 & 3 & \textbf{19.67} & 2 \\
\addlinespace
W1 & 12.53 & 4 & 8.03 & 5 \\
W2 & \textbf{16.33} & 1 & 8028 & 6 \\
W3 & 10.12 & 6 & \textbf{22.11} & 1 \\
\addlinespace
S1 & 10.27 & 5 & 6.42 & 7 \\
\bottomrule
\end{tabular}
\end{table}


When perturbation-definition templates are added to the prompt, both generation modes become more stable overall, but the degree of improvement differs. No-CoT benefits more consistently, especially on MHPP, while CoT experiences comparatively smaller changes. This suggests that template-based task clarification primarily resolves ambiguity in direct code generation, but does not fully mitigate the points at which perturbations influence CoT's multi-step reasoning.

Taken together, these findings indicate that robustness differences between CoT and No-CoT do not manifest as a uniform advantage for either mode. Instead, each generation strategy has characteristic vulnerability patterns, shaped by how it processes and interprets the input prompt. This underscores that CoT and No-CoT should be evaluated as distinct robustness behaviors, rather than as alternatives where one universally outperforms the other.

We further examine how robustness evolves as the sampling budget increases by checking whether $|RD|$ grows monotonically across $k \in \{1,5,10\}$ within each (model, dataset, temperature, perturbation, mode) configuration. When aggregating results by perturbation type, the number of monotonic-increase cases ranges from 0/8 to 2/8 per perturbation and per generation mode. This distribution indicates that no perturbation family inherently leads to consistently increasing degradation as $k$ grows; rather, for all perturbations, most configurations retain the possibility of recovery when more samples are drawn.

To statistically examine whether the degradation differences between CoT and No-CoT are significant, we conduct Wilcoxon signed-rank tests over paired RD values across perturbation configurations. Table~\ref{tab:rq2_wilcoxon} summarizes the number of paired observations, test statistics, p-values, and effect sizes.


\begin{table}[t]
\centering
\caption{Wilcoxon signed-rank test on relative degradation (RD) between CoT and No-CoT across datasets and models. Effect size (Eff.) is negligible (N), small (S), medium (M), or large (L).}
\label{tab:rq2_wilcoxon}
\footnotesize
\setlength{\tabcolsep}{3pt}
\begin{tabular}{@{}llccccc@{}}
\toprule
\textbf{Model} & \textbf{Dataset} & \textbf{$p$} & \textbf{$r$} & \textbf{Eff.} \\
\midrule
DeepSeek & MHPP  & $5.03\times10^{-4}$  & 0.543 & L \\
Qwen     & MHPP  & $4.42\times10^{-1}$  & 0.121 & S \\
DeepSeek & BCB   & $9.86\times10^{-6}$  & 0.682 & L \\
Qwen     & BCB   & $3.18\times10^{-11}$ & 0.848 & L \\
\bottomrule
\end{tabular}
\end{table}

However, on BigCodeBench the difference becomes highly significant for Qwen-Coder ($p<10^{-10}$), accompanied by a very large effect size. These mixed outcomes indicate that the robustness relationship between CoT and No-CoT is not consistent across models or evaluation benchmarks. Instead, the effect of prompting mode depends strongly on the interaction between the model architecture and the perturbation characteristics of the dataset.


\begin{findingbox}{Hypothesis Decision (RQ2)}
We reject $H_0^{2}$ for several perturbation families, though not uniformly across all conditions. These results indicate that robustness differences depend on perturbation type rather than exhibiting complete equivalence.
\end{findingbox}

\subsection{\emph{\rqthree}}

We evaluate $H_0^{3}$ by examining whether early-stage uncertainty measures are statistically associated with final code-generation failure using correlation and AUROC analyses.

To statistically assess this relationship, we compute Spearman rank correlations between early uncertainty metrics and the binary failure outcome. 
Spearman's $\rho$ measures the strength of monotonic association, while the corresponding $p$-value evaluates whether the observed correlation differs significantly from zero. 
A statistically significant correlation ($p < 0.05$) would indicate that early uncertainty provides evidence for predicting generation failure.
To assess whether early uncertainty can anticipate final pass@k outcomes, we analyze entropy and probability differential over the first 30--40\% of generated tokens across MHPP and BigCodeBench, under both CoT and No-CoT settings. Table~\ref{tab:uncertainty_correlation} summarizes the relationship between early-stage uncertainty and final generation outcomes.


Overall, early uncertainty exhibits a consistent but weak association with failure. 
Across all models, the magnitude of the Spearman correlations remains small ($|\rho| \approx 0.1$--$0.2$), indicating only a weak monotonic relationship between early uncertainty and final outcomes. 
Moreover, most correlations are not statistically significant ($p > 0.05$), suggesting that early uncertainty signals alone do not provide strong statistical evidence for predicting failure. 
The only marginally significant case appears for GPT-OSS-20B ($\rho \approx 0.19$, $p \approx 0.07$), which approaches but does not reach conventional significance thresholds.

At the level of individual uncertainty metrics, the discriminative power of early uncertainty is close to chance. As shown in Table~\ref{tab:uncertainty_correlation}, AUROC values based on entropy or probability differential alone range from approximately 0.55 to 0.60 across models. These results indicate that while early uncertainty is directionally aligned with failure, no single uncertainty measure is sufficient to reliably separate successful from failed generations.

Model-level differences are therefore quantitative rather than qualitative. DeepSeek exhibits slightly stronger correlations and higher AUROC values than Qwen and GPT, suggesting that its failures are more often preceded by detectable early instability. In contrast, Qwen tends to produce stable, low-entropy early tokens even for incorrect solutions, with errors more likely to emerge later during logical transitions or code synthesis. As a result, early uncertainty is less informative for Qwen, though the overall trend remains consistent across models.

Importantly, these findings are robust to how early uncertainty is measured. Restricting analysis to early code tokens (CoT--code-only) or including the entire early window with natural-language reasoning (CoT--full) yields nearly identical correlation patterns. Likewise, CoT and No-CoT exhibit comparable early-uncertainty behavior, indicating that the presence of a reasoning chain does not fundamentally alter how early uncertainty aligns with final success or failure.

Finally, while individual uncertainty metrics offer limited predictive value, aggregating multiple early uncertainty features using a lightweight logistic regression substantially improves separability, achieving AUROC values in the range of 0.66--0.69 across models. This suggests that early uncertainty should be interpreted not as a standalone predictor, but as a weak diagnostic signal whose value emerges through structured aggregation.


\begin{table}[!ht]
\centering
\caption{Early-stage uncertainty metrics as predictors of generation failure. 
Spearman's $\rho$ indicates correlation with failure, and $p$-values report the statistical significance of the correlation. 
AUROC measures predictive accuracy.}
\label{tab:uncertainty_correlation}

\begin{tabular}{@{}lcccc@{}}
\toprule
 & \multicolumn{2}{c}{\textbf{Entropy}} & \multicolumn{2}{c}{\textbf{Probability Difference}} \\
\cmidrule(lr){2-3}\cmidrule(lr){4-5}
\textbf{Model} & $\rho$ & $p-value$ & $\rho$ & $p-value$ \\
\midrule
DeepSeek-Coder   & $-0.08$ & $0.34$ & $+0.04$ & $0.45$ \\
Qwen2.5-Coder    & $-0.04$ & $0.45$ & $+0.04$ & $0.45$ \\
GPT-OSS-20B      & $-0.19$ & $0.07$ & $+0.19$ & $0.07$ \\
\addlinespace
\multicolumn{5}{c}{\textbf{Failure Prediction (AUROC)}} \\
\cmidrule{1-5}
\textbf{Model} & \multicolumn{2}{c}{Entropy} & \multicolumn{2}{c}{Probability Difference} \\
\midrule
DeepSeek-Coder   & \multicolumn{2}{c}{0.43} & \multicolumn{2}{c}{0.46} \\
Qwen2.5-Coder    & \multicolumn{2}{c}{0.48} & \multicolumn{2}{c}{0.48} \\
GPT-OSS-20B      & \multicolumn{2}{c}{0.43} & \multicolumn{2}{c}{0.43} \\
\bottomrule
\end{tabular}
\end{table}

\begin{findingbox}{Hypothesis Decision (RQ3)}
We reject $H_0^{3}$; however, the predictive strength of individual uncertainty metrics remains limited. This suggests that early uncertainty functions as a weak diagnostic signal rather than a strong standalone predictor.
\end{findingbox}

\subsection{\emph{\rqfour}}


We evaluate $H_0^{4}$ by analyzing whether perturbations systematically alter CoT reasoning trajectories relative to structural anchors, using anchor-aligned spike distributions and deformation classification.

Although RQ3 shows that early-stage uncertainty is not a reliable predictor of final correctness, especially for models like Qwen, it remains useful for analyzing how perturbations change the generation process under CoT. In RQ4, we therefore treat uncertainty as a diagnostic signal rather than a decision criterion: the location of early uncertainty spikes indicates where the trajectory starts to deviate. These spikes do not deterministically fix the final outcome, but they consistently align with points where reasoning, symbol grounding, or control flow become unstable.

Our analysis shows that perturbations reshape the entire CoT trajectory rather than only affecting the natural-language prefix. Under different perturbation families and temperatures, we observe systematic deformation patterns—lengthening, branching, and simplification—that influence the resulting program. These deformations emerge around a small set of structural anchors connecting reasoning and code, rather than from isolated token-level noise.

\subsubsection{Anchor-Aligned Localization of Uncertainty}
\label{sec:anchor}


To characterize where uncertainty first arises along the generation trajectory, we begin by examining the position of the first major uncertainty spike in absolute (token-scale) terms.

We further investigate whether these spikes systematically align with structurally meaningful positions along the reasoning-to-code trajectory.

To evaluate whether the proposed anchors capture meaningful structural commitment points, we test whether linguistic perturbations significantly shift spike localization relative to structural anchors in CoT trajectories ($H_0^{4a}$). 
If anchors correspond to structurally sensitive stages of reasoning, spike positions should remain stable relative to anchors even when perturbations alter surface-level prompt wording.

We therefore compare anchor-aligned spike distances $\Delta = (S - A_k)/T$ between clean and perturbed prompts using the Wilcoxon signed-rank test for paired samples, since the same tasks are evaluated under both prompt conditions. 
As a robustness check, we additionally apply a two-sample Kolmogorov--Smirnov (KS) test to compare the full distributions of spike distances between the two prompt conditions.

Across perturbation families, the Wilcoxon signed-rank tests (Table~\ref{tab:rq4_h4a}) do not reveal statistically significant differences in anchor-aligned spike localization ($p > 0.05$ for all perturbation types). Effect sizes are consistently small ($r < 0.14$), suggesting that perturbations introduce only minor changes in spike positions relative to structural anchors. 
Similarly, the KS tests comparing the full distributions of spike distances between clean and perturbed prompts yield small KS statistics ($D \leq 0.114$) and non-significant $p$-values. 

\begin{table}[t]
\centering
\caption{Statistical results for $H_0^{4a}$ across perturbation families.}
\begin{tabular}{lcccc}
\toprule
Perturbation & Wilcoxon $p$ & Effect size $r$ & KS $D$ & KS $p$ \\
\midrule
C1 & 0.0548 & 0.133 & 0.114 & 0.1288 \\
C2 & 0.1884 & 0.091 & 0.0667 & 0.7404 \\
C3 & 0.8163 & 0.016 & 0.0667 & 0.7404 \\
S1 & 0.6386 & 0.032 & 0.0810 & 0.4979 \\
W1 & 0.7284 & 0.024 & 0.0571 & 0.8839 \\
W2 & 0.6682 & 0.030 & 0.0429 & 0.9908 \\
W3 & 0.5649 & 0.040 & 0.0619 & 0.8171 \\
\bottomrule
\end{tabular}
\label{tab:rq4_h4a}
\end{table}

Overall, the results do not provide sufficient evidence to reject $H_0^{4a}$. This indicates that linguistic perturbations do not substantially alter the localization of uncertainty spikes relative to structural anchors. 
The stability of spike localization across perturbation conditions supports the validity of the proposed anchors as structurally meaningful reference points for analyzing reasoning trajectories.

Having established the robustness of anchor-aligned spike localization, we next examine how uncertainty spikes are distributed along the generation trajectory.

\begin{figure*}[!ht]
  \centering
  \includegraphics[width=\textwidth]{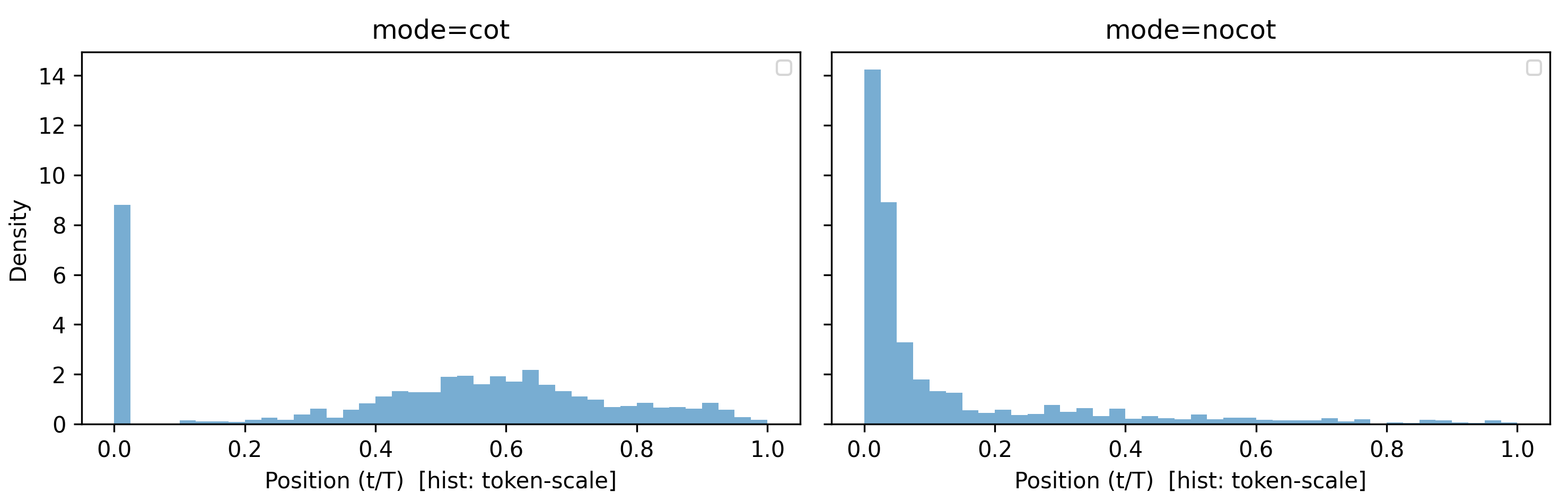}
  \caption{
  \textbf{Distribution of the first uncertainty spike position (token scale).} The x-axis denotes the normalized position of the first major uncertainty spike $S/T$ along the generated trajectory, aggregated across all models, datasets, temperatures, and perturbation conditions. Under CoT, spikes are broadly distributed across the trajectory with a long tail toward later tokens, whereas under No-CoT they are heavily front-loaded in the very early stage of generation. This indicates that CoT tends to redistribute uncertainty toward later stages of reasoning, while No-CoT typically exposes instability almost immediately after
  code generation begins.
  }
  \label{fig:first_spike_position}
\end{figure*}

Figure~\ref{fig:first_spike_position} characterizes the overall positional distribution of the first uncertainty spike (normalized by total token length), pooling together results across all models, datasets, temperatures, and perturbation conditions. We observe that No-CoT trajectories are highly front-loaded: most spikes occur within the first 10\% of tokens, indicating that instability typically emerges very early once the model begins committing to executable structure. In contrast, CoT exhibits a much broader distribution, with a substantial fraction of spikes appearing in the middle or later stages of generation. This pattern suggests that CoT does not eliminate uncertainty but redistributes it along the trajectory, delaying or diffusing where instability first manifests.

\begin{figure*}[!ht]
  \centering
  \includegraphics[width=\textwidth]{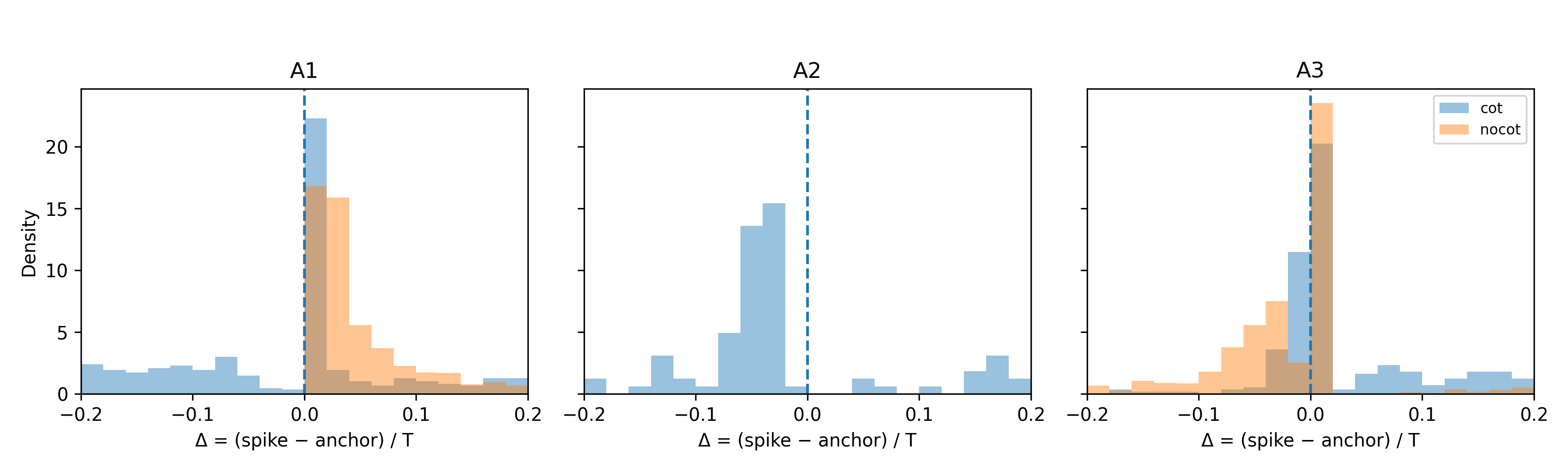}
  \caption{
  \textbf{Alignment between the first uncertainty spike and structural anchors.}
  Each panel shows the distribution of normalized distances 
  $\Delta = (S - A_k)/T$ between the first uncertainty spike $S$ and a structural anchor $A_k$.
  Vertical dashed lines indicate the anchor location ($\Delta = 0$).
  Negative values correspond to spikes occurring before the anchor, and positive values to spikes occurring after.
  Under CoT, uncertainty spikes concentrate near the reasoning--code transition (A1) and algorithmic articulation (A3),
  whereas no-CoT trajectories exhibit weaker and less structured alignment.
  }
  \label{fig:spike_anchor_alignment}
\end{figure*}

We next analyze how these spikes are positioned relative to structural anchors that connect reasoning and code (Figure~\ref{fig:spike_anchor_alignment}).

Figure~\ref{fig:spike_anchor_alignment} reports the distribution of the normalized distance $\Delta = (S - A_k)/T$ between the first uncertainty spike $S$ and each anchor $A_k$. Across models and perturbation settings, uncertainty spikes are not evenly distributed along the trajectory. Under CoT, spikes concentrate around a small number of structurally sensitive anchors, most notably the reasoning--code transition (A1) and the onset of algorithmic articulation (A3). This concentration indicates that perturbations tend to destabilize the generation process when the model commits to executable structure rather than during early surface-level reasoning. In contrast, no-CoT trajectories exhibit weaker and less consistent alignment, suggesting that deviation emerges earlier and in a more diffuse manner.

\subsubsection{Observed Structural Deformation Patterns}

Empirically, perturbations modify the generation process in consistent ways across models and datasets. Instead of only increasing token-level variance, they change the global structure of the reasoning-to-code pathway. Three deformation patterns appear frequently, summarized in Table~\ref{tab: Trajectory}.

\begin{table*}[!ht]
\centering
\caption{Trajectory deformation patterns induced by prompt perturbations in CoT reasoning.}
\textit{Note:} Patterns are identified through qualitative analysis of reasoning trajectories across models and benchmarks.
\label{tab:trajectory_patterns}
\begin{tabular}{@{}lp{8cm}lp{4cm}@{}}
\toprule
\multicolumn{1}{c}{\textbf{Pattern}} & \textbf{Description} & \textbf{Typical Triggers} & \textbf{Effect on Pass@$k$} \\
\midrule
\textbf{Lengthening} & Reasoning becomes longer, with redundant or circular steps; code generation is delayed. & W2, S1 at high $T$ & Pass → Fail \\
\addlinespace
\textbf{Branching} & Early uncertainty causes multiple competing interpretations; reasoning drifts between inconsistent pathways. & C2, C3 at high $T$ & Strongly harmful (Pass → Fail) \\
\addlinespace
\textbf{Simplification} & Reasoning collapses into fewer steps; model jumps earlier into code. & W1, W2 at low $T$ & Mixed (Pass $\leftrightarrow$ Fail) \\
\bottomrule
\end{tabular}
\label{tab: Trajectory}
\end{table*}

To examine whether these qualitative patterns correspond to systematic structural instabilities along the generation trajectory,
we further analyze their association with structural anchors.

\begin{figure}[t]
  \centering
  \includegraphics[width=0.5\textwidth]{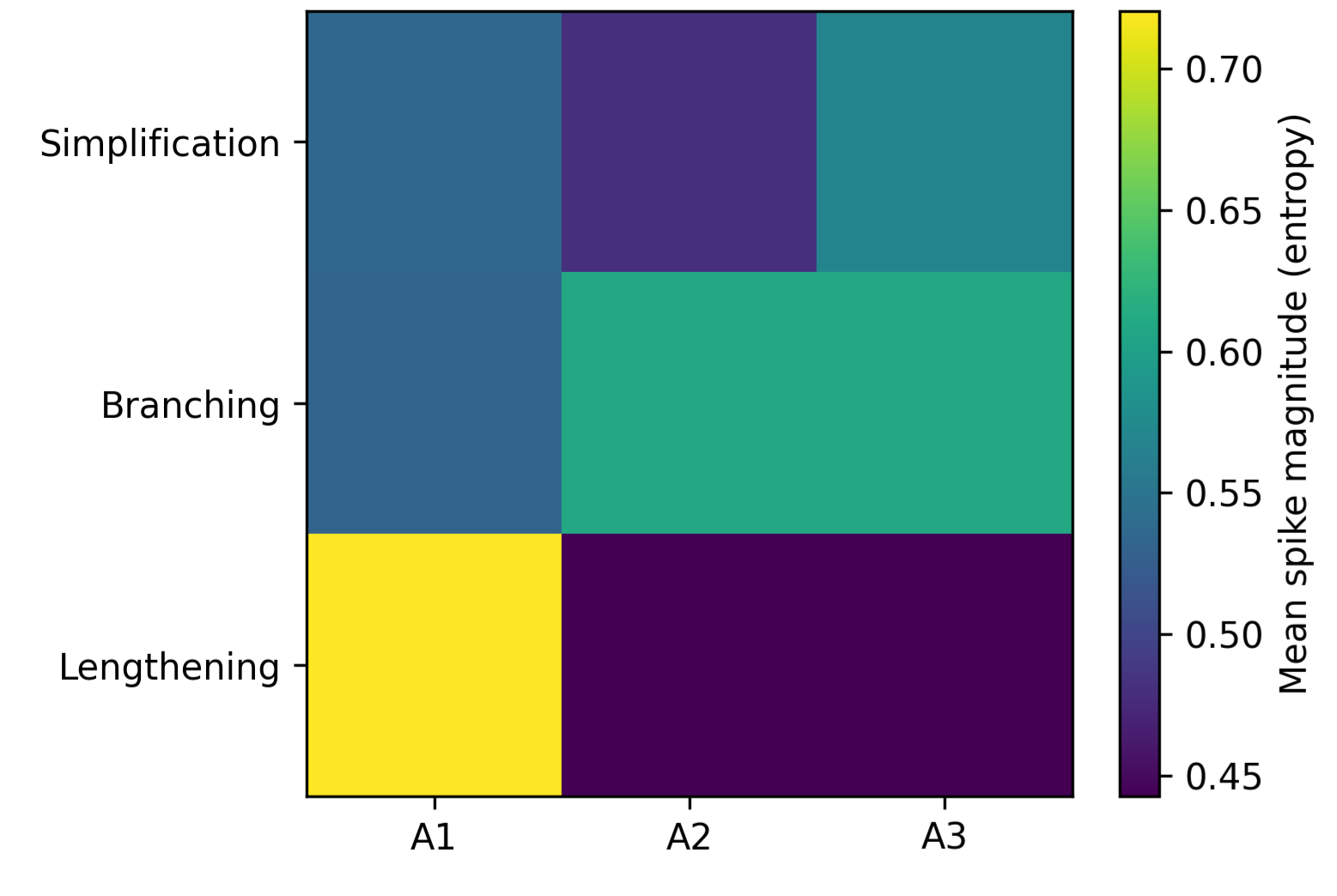}
  \caption{
  \textbf{Association between structural anchors and deformation patterns under CoT.}
  Each cell reports the mean magnitude of the first uncertainty spike
  associated with a given deformation pattern at a specific anchor.
  Higher values indicate stronger uncertainty concentration.
  Lengthening is most strongly associated with early instability near the reasoning--code transition (A1),
  while branching shows elevated uncertainty around symbol grounding and algorithmic articulation (A2/A3).
  Simplification exhibits weaker and more diffuse anchor association, consistent with early commitment
  and reduced trajectory complexity.
  }
  \label{fig:anchor_deformation}
\end{figure}


Figure~\ref{fig:anchor_deformation} provides quantitative evidence that the observed deformation patterns are not uniformly distributed along the generation trajectory, but are instead anchored at specific structurally sensitive positions. Lengthening is primarily associated with increased uncertainty near the reasoning--code transition (A1), indicating delayed commitment to executable structure. Branching exhibits stronger uncertainty around symbol grounding and early algorithmic articulation (A2/A3),
reflecting competing internal representations before control flow stabilizes. In contrast, simplification shows comparatively weaker and more diffuse anchor alignment, consistent with early commitment to a single solution pathway.

Lengthening increases the amount of natural-language reasoning before the first line of code. Longer chains expose more positions where perturbation-induced uncertainty can alter intermediate assumptions. As the model rarely revisits earlier steps, small deviations introduced in the middle of the chain tend to accumulate and carry over into the implementation. This error-accumulation effect makes lengthened trajectories more likely to turn previously correct cases into failures.

Branching is the most disruptive deformation. It corresponds to trajectories in which the model temporarily maintains multiple competing hypotheses about the task. Stronger character-level perturbations (C2/C3) often shift or weaken key lexical cues, leading to segments of reasoning that encode inconsistent algorithmic assumptions. When code generation starts, the model commits to a mixture of these assumptions, producing mismatched variable usage, contradictory control flow, or inconsistent edge-case handling. Compared to lengthening, branching reflects structural inconsistency in the inferred specification rather than gradual drift, and is therefore associated with larger performance drops.

Simplification shortens the CoT chain and leads the model to commit earlier to a single solution pattern. Under mild word-level perturbations at lower temperature, shorter chains reduce the number of positions where uncertainty spikes can redirect the reasoning. When the model already has a strong prior for the correct algorithm, early commitment avoids the detours and alternative branches that longer CoT sometimes introduces, which explains the Fail→Pass cases we observe. However, if the early interpretation is incorrect, simplification simply fixes an erroneous plan and can still lead to Pass→Fail; its impact is therefore asymmetric and context-dependent.

\subsubsection{Uncertainty Signatures and Deformation Modes}

Finally, we link the deformation patterns in Table~\ref{tab: Trajectory} with the uncertainty behavior at these anchors. Table~\ref{tab:uncertainty_anchor} summarizes the relationship between the location and shape of the first major uncertainty spike, the dominant anchor it affects, the resulting deformation type, and the typical outcome.


\begin{table*}[!ht]
\centering
\caption{Uncertainty signatures, structural anchors, and resulting deformation outcomes.}
\label{tab:uncertainty_anchor}
\begin{tabular}{@{}lp{7.5cm}lp{3cm}@{}}
\toprule
\multicolumn{1}{c}{\textbf{Uncertainty Signature}} & \textbf{Anchor Location} & \textbf{Deformation} & \textbf{Outcome} \\
\midrule
\textbf{Early, sharp spike} & Reasoning--code transition or symbol grounding (A1/A2) & Branching & Strong Pass → Fail \\
\addlinespace
\textbf{Mid-trajectory, gradual rise} & CoT elaboration phase & Lengthening & Moderate Pass → Fail \\
\addlinespace
\textbf{Flat or suppressed profile} & Early reasoning (stable interpretation) & Simplification & Mixed (Pass $\leftrightarrow$ Fail) \\
\bottomrule
\end{tabular}
\end{table*}

Early, sharp spikes near the reasoning–code transition or symbol-grounding anchor are commonly followed by branching trajectories and strong Pass→Fail degradation. Mid-trajectory, gradual increases in uncertainty tend to co-occur with lengthened reasoning, consistent with error accumulation across extended chains. Flat or suppressed uncertainty in the early reasoning phase is associated with simplified trajectories: when the initial interpretation is correct, this leads to Fail→Pass recovery; when it is wrong, the same mechanism produces premature Pass→Fail.  

Overall, RQ4 shows that perturbations reshape CoT generation by interacting with a small set of structural anchors, and that the resulting deformation patterns are reflected in the uncertainty signatures at those anchors. These mechanisms explain why CoT does not yield a uniform robustness advantage, and why different perturbation families exhibit distinct robustness profiles under CoT and No-CoT.

To evaluate $H_0^{4b}$, we examine whether the type of reasoning trajectory deformation is statistically associated with the final generation outcome (pass or fail). 
We apply a chi-square test of independence to evaluate whether deformation patterns are statistically associated with generation success or failure. 
Effect sizes are reported using Cramér's $V$, which quantifies the strength of association between the two categorical variables.

The results show a statistically significant association between deformation pattern and generation outcome ($\chi^2 = 25.93$, $p = 2.34 \times 10^{-6}$). The effect size measured by Cramér's V is $V = 0.094$, indicating a small but statistically reliable association.
Although the magnitude of the effect is modest, this is expected in large-scale generative systems where final correctness is influenced by multiple interacting factors beyond trajectory structure alone (e.g., task difficulty, model capability, and prompt formulation). 
Given the large sample size and the consistent direction of the association across perturbation conditions, the result nevertheless provides reliable evidence that certain trajectory deformation patterns are systematically linked to generation failures. 
This finding is consistent with our trajectory-based interpretation, in which deformation reflects localized reasoning instability rather than being the sole determinant of generation failure.
These results suggest that certain trajectory deformation patterns are systematically linked to generation failures, supporting the interpretation that instability in the reasoning trajectory reflects underlying reasoning breakdown.



\begin{findingbox}{Hypothesis Decisions (RQ4)}
We cannot reject $H_0^{4a}$ but reject $H_0^{4b}$. 
The stability of spike localization relative to structural anchors indicates that perturbations do not substantially shift where uncertainty emerges along the reasoning trajectory. 
However, trajectory deformation patterns are significantly associated with generation outcomes, suggesting that failures arise not from relocating uncertainty spikes but from how reasoning trajectories deform around structurally sensitive anchors.
\end{findingbox}

%% file: Discuss.tex
\section{Discussion}
\label{sec:discussion}
Section~\ref{sec:results} formally evaluated $H_0^{1}$–$H_0^{4}$ and reported explicit hypothesis decisions for each research question. We now interpret these decisions to refine the theoretical understanding of CoT in code generation and articulate their broader implications.

\subsection{RQ1: Contingent Performance Effects}

The rejection of $H_0^{1}$ refines the prevailing consensus that CoT aids complex reasoning. While our results confirm that CoT can improve performance on natural-language-intensive tasks (e.g., DeepSeek on BigCodeBench) or when prompts lack explicit structure (e.g., Qwen on standard MHPP), this advantage is not guaranteed. It dissolves under two conditions: (1) for models like CodeLlama, which lack reasoning supervision and treat CoT as noisy context; and (2) when prompts are made semantically precise, as adding perturbation definitions erased Qwen’s CoT edge.

These findings suggest CoT’s primary utility may be ambiguity resolution—a benefit that diminishes when the task specification is already structurally explicit. Rather than being inherently beneficial, CoT’s effectiveness depends on model supervision and task formulation.

\subsection{RQ2: Distinct Robustness Profiles}

Rejecting $H_0^{2}$ clarifies that robustness effects cannot be reduced to a simple ranking between prompting modes. A common intuition is that longer reasoning chains might be more vulnerable to perturbations; however, our results show the relationship is more nuanced.

CoT and No-CoT exhibit divergent, not ranked, vulnerability profiles. No-CoT failures concentrate in lexical–semantic misalignment (e.g., altered key verbs), while CoT failures more often arise from narrative–structural disruption (e.g., broken grammatical coherence). Thus, robustness is not a property conferred by reasoning length alone, but a reflection of how each generation mode parses—and can be misled by—linguistic variation.

\subsection{RQ3: Uncertainty as a Local Instability Signal}

The rejection of $H_0^{3}$ demonstrates that early-stage uncertainty is associated with downstream failure. However, the modest effect sizes observed indicate that uncertainty is better interpreted as a localized instability signal rather than a reliable standalone predictor.

Uncertainty spikes tend to localize at structurally meaningful commitment points in the reasoning trajectory. For models that permit branching, such spikes often precede failure. For models that suppress branching, errors may arise from subtle, low-uncertainty mis-specifications at these same anchors. Thus, uncertainty highlights instability at critical structural junctures but does not deterministically determine final outcomes.

\subsection{RQ4: Anchor-Mediated Trajectory Deformation}

Rejecting $H_0^{4}$ supports the anchor-deformation framework as a mechanistic explanation for CoT’s contingent behavior. Perturbations do not uniformly influence the reasoning trajectory; instead, they act upon structural anchors—pivotal points where the model commits to organizing and implementing the task.

The primary anchors include the reasoning–code transition, symbol grounding, and algorithmic articulation. Minor linguistic shifts at these anchors deform trajectories in systematic ways:

\begin{itemize}
\item \textbf{Lengthening:} Hesitation at transition anchors produces redundant elaboration.
\item \textbf{Branching:} Inconsistent commitments at grounding or algorithmic anchors spawn parallel, incompatible reasoning threads.
\item \textbf{Simplification:} Overconfident early commitment collapses reasoning into a rigid solution template.
\end{itemize}

These deformation modes explain the heterogeneous performance transitions observed under perturbation. Anchors act as leverage points; perturbations destabilize them; and the resulting structural deformation determines eventual success or failure.






\subsection{Synthesis: The Anchor--Deformation Theory of CoT in Code Generation}
Collectively, rejecting $H_0^{1}$–$H_0^{4}$ supports a structural reinterpretation
of CoT reasoning in LLM-based code generation. Rather than
functioning as a universally beneficial reasoning enhancement, CoT acts as a
structural transformation of the generation trajectory. In particular, CoT
introduces explicit intermediate commitments—referred to here as
\emph{structural anchors}—that mediate how models interpret, organize, and
implement task specifications during generation.
\begin{theorybox}
\textbf{Anchor--Deformation Theory of CoT.}\\
CoT reasoning structures generation trajectories around
commitment points (\emph{anchors}). The stability of these anchors governs
correctness and robustness, while perturbations that destabilize anchors can
induce \emph{anchor-localized trajectory deformations} that propagate through
subsequent reasoning and generation steps.
\end{theorybox}
These anchors correspond to moments in the reasoning trajectory where the
model commits to key aspects of the task representation, such as problem
interpretation, symbol grounding, or algorithmic formulation. From this
perspective, the empirical phenomena observed in our study are not independent
effects. The contingent performance impact of CoT (RQ1), the divergent
robustness profiles of CoT and No-CoT (RQ2), the localization of uncertainty
signals (RQ3), and the deformation of reasoning trajectories under perturbation
(RQ4) all emerge from a shared underlying mechanism: the stability or
instability of anchor commitments during generation.
Evidence from our empirical analyses (Section~\ref{sec:results}) indicates
that uncertainty spikes frequently occur near these anchors, marking points
where the model struggles to stabilize its internal representation of the
task. When anchors remain stable, reasoning trajectories tend to maintain
semantic alignment with the task specification, producing correct and
interpretable implementations. However, when anchors are destabilized—through
lexical noise, grammatical shifts, or subtle semantic variation—perturbations
can induce structural deformations of the reasoning trajectory. These
deformations propagate downstream, amplifying early misinterpretations and
ultimately leading to incorrect or insecure code.
This perspective reframes how CoT should be evaluated in LLM4Code systems.
Rather than assessing CoT solely through aggregate accuracy metrics,
evaluation should consider the structural stability of reasoning commitments
and how perturbations influence anchor-localized trajectory dynamics. By
focusing on the stability of structural anchors, researchers can better
understand when CoT improves reliability and when it introduces new failure
modes.

\subsection{Robustness of Security Alignment}

\begin{table}[t]
\centering
\caption{Security alignment results under different perturbations (Semgrep). Findings denote blocking violations detected across 56 rules.}
\label{tab:security_semgrep}
\renewcommand{\arraystretch}{1.15}
\setlength{\tabcolsep}{5pt}
\begin{tabular}{l|cc|cc}
\hline
 & \multicolumn{2}{c|}{\textbf{CoT}} & \multicolumn{2}{c}{\textbf{No-CoT}} \\
\textbf{Perturbation} 
 & Findings & Targets & Findings & Targets \\
\hline
Nomin & 0 & 2051 & 2 & 2100 \\
C1    & 4 & 2070 & 3 & 2100 \\
C2    & 6 & 2100 & 0 & 2100 \\
C3    & 4 & 2059 & 0 & 2100 \\
W1    & 0 & 2100 & 4 & 2100 \\
W2    & 2 & 2100 & 0 & 2100 \\
W3    & 10 & 2100 & 0 & 2100 \\
S1    & 0 & 2100 & 4 & 2100 \\
\hline
\end{tabular}
\end{table}

We further evaluated security alignment using automated static analysis (Semgrep) across 56 security rules, scanning approximately 2,100 targets per group with a parsing success rate of 98.2\%.

As shown in Table~\ref{tab:security_semgrep}, the results reveal that CoT functions as a double-edged mechanism under perturbations.

\textbf{Semantic Resilience.} Under high-level structural perturbations (e.g., back-translation and synonym insertion), CoT demonstrates greater robustness. The reasoning process appears to denoise complex prompts, preserving security constraints that the No-CoT baseline fails to maintain.

\textbf{Lexical Fragility.} Under character-level noise and grammatical variation, CoT becomes highly sensitive. Minor lexical shifts can induce reasoning drift, causing the model to bypass security boundaries while still generating syntactically valid code.

\textbf{Utility–Security Trade-off.} The absence of findings in No-CoT under heavy noise likely reflects failure to generate complex logic. CoT’s persistence in producing functional code from noisy inputs increases expressive capability but inadvertently expands the attack surface.

These findings extend the anchor-deformation perspective: semantic robustness emerges when anchors remain stable under paraphrasing, whereas lexical fragility arises when subtle perturbations destabilize structural commitments.

%% file: Implication.tex
\section{Implications} \label{sec:implication}

Our findings suggest that the implications of CoT reasoning in code generation are best understood through three practical observations. First, CoT should not be treated as a universal default for all coding tasks. Second, its value appears to be most pronounced in ambiguity-heavy settings, where the model must interpret, organize, and reconcile complex or noisy instructions. Third, when CoT is deployed, its reliability should be assessed not only through final code correctness but also through anchor-aware monitoring of the generation trajectory. Together, these implications shift the discussion from whether CoT helps in general to when it is useful and how it can be deployed more reliably.

\subsection{Not All Tasks Should Use CoT by Default}

A first implication of our study is that CoT should not be assumed to improve code generation uniformly across all tasks. For relatively simple, well-specified, and highly deterministic problems, direct generation or only light planning may provide a more stable and efficient solution. In such cases, forcing the model to externalize extended reasoning may introduce unnecessary variance, increase the opportunity for structural drift, and provide limited functional benefit.

This suggests that future evaluation and system design should move beyond a one-size-fits-all view of reasoning. Rather than asking whether CoT is generally better, a more useful question is under what task conditions it becomes helpful or harmful. Benchmark design should therefore distinguish between tasks that are inherently straightforward and tasks that involve ambiguity, underspecification, or structural noise. In practical settings, the decision to enable CoT should be validated against the characteristics of the target workload rather than adopted as a default prompting strategy.

More broadly, our results support a selective view of reasoning: explicit step-by-step generation is not a universal marker of stronger performance, but a context-dependent strategy whose benefits must be weighed against its potential instability.

\subsection{The Real Value of CoT Emerges in Ambiguity-Heavy Settings}

A second implication is that the practical value of CoT appears to concentrate in ambiguity-heavy settings. These include tasks with long or complex instructions, realistic software engineering scenarios, incomplete or noisy specifications, and prompts containing structural perturbations. Under such conditions, the model must do more than retrieve a familiar solution pattern; it must interpret, organize, and reconcile multiple constraints over a longer generation trajectory. This is where explicit reasoning may become genuinely useful.

Our findings therefore point toward a more targeted role for CoT. Its strength may lie less in routine deterministic problems and more in situations where the model benefits from explicitly maintaining intermediate structure before committing to code. This has implications for both evaluation and deployment. Evaluation should better distinguish between clean benchmark settings and more realistic ambiguity-heavy scenarios. Likewise, code-generation systems may benefit from adaptive strategies that reserve CoT for tasks exhibiting high ambiguity, complex instruction structure, or elevated uncertainty.

In this sense, the contribution of our work is not simply to question whether CoT helps, but to clarify where its value is most likely to materialize. The answer appears to depend not only on model capability, but also on the structural demands of the task itself.

\subsection{Deploying CoT Requires Anchor-Aware Monitoring}

A third implication is that, when CoT is deployed, it should not be evaluated solely through final code correctness. Pass/fail outcomes remain important, but they provide only an endpoint judgment and reveal little about how the generation process became unstable. Our results show that failure often emerges through localized deformation around key structural anchors, suggesting that CoT systems should be monitored at the trajectory level rather than only at the final-output level.

This motivates an anchor-aware monitoring perspective. In practice, systems using CoT should pay particular attention to abnormal uncertainty or instability near critical transition points such as the \textit{reasoning-to-code transition}, \textit{symbol commitment}, and \textit{algorithmic articulation}. These anchors appear to concentrate moments where the model moves from abstract interpretation to concrete implementation, and thus where latent reasoning instability may become operationally consequential.

Such monitoring could support several practical mechanisms. For example, unusual uncertainty spikes around anchor regions could trigger fallback strategies, request clarification, simplify the reasoning path, or route the case for human review. More broadly, these signals offer a richer diagnostic lens for studying and managing reasoning robustness. In this way, anchor-aware monitoring transforms reasoning analysis from a post hoc interpretive tool into a potentially actionable mechanism for improving reliability.

\subsection{From Capability to Selective and Reliable Use}

These implications point to a shift in how CoT should be understood in code generation. The key question is no longer whether models can produce step-by-step reasoning, but when such reasoning should be used, where it adds genuine value, and how its instability can be detected before it leads to failure. CoT should not be treated as a universal default; it appears most useful in ambiguity-heavy settings, and its deployment should be accompanied by anchor-aware monitoring of structurally sensitive stages.

This reframing moves the discussion from capability to reliability. The practical future of CoT in code generation will depend less on its best-case gains and more on whether it can be selectively applied and diagnostically monitored under realistic conditions. From this perspective, robust code-generation systems are not those that always reason more, but those that reason when needed and remain observable when reasoning begins to drift.

%% file: LimitationAndFutureWork.tex
\section{Limitation And Future Works}\label{sec:limi}

Our study focuses on open-source models (up to 13B parameters) and linguistic perturbations. Larger, proprietary, or differently trained models (e.g., with extensive RL alignment) may exhibit different anchor dynamics, which merits verification. Furthermore, our perturbation set is linguistic; future work should investigate program-aware perturbations (e.g., to API names, type signatures, or code examples within the prompt), which may interact with reasoning anchors in distinct ways.

These observations point toward several concrete pathways for improving future model design:
\begin{itemize}
\item Anchor-Aware Training: Incorporate adversarial examples that perturb structural anchors to teach models to maintain coherence.
\item Trajectory Regularization: Develop decoding-time methods to penalize excessive lengthening or prune branching inconsistencies.
\item Modular Interfaces: Architectures that separate planning from synthesis could harden the critical reasoning–code transition.
\item Prompt-Adaptive Strategy Selection: Our discovery of complementary failure profiles suggests training a \textit{decision model} to analyze the input prompt and dynamically select between CoT and direct generation. Such a model could be trained on data where the optimal strategy is known, potentially using features like prompt complexity, lexical variability, and estimated ambiguity.
\end{itemize}

Finally, while we identify structural anchors diagnostically, their real-time identification and protection remain an open engineering challenge. Translating our offline analysis into runtime monitoring systems that can detect anchor instability and trigger corrective actions (e.g., localized regeneration, confidence-based fallback) is a crucial next step toward deploying robust, reasoning-aware code generators.

%% file: ThreatstoValidity.tex
\section{Threats to Validity} \label{sec:threats}
We discuss threats to the validity of our study following established categories in empirical software engineering.

\subsection{Internal Validity}
Internal validity concerns whether the observed effects can be causally attributed to the manipulated factors.

Perturbation Implementation: Our perturbations, while designed to mimic real-world variations, are generated automatically and may not fully capture the nuanced ways humans rephrase requests. However, we employed multiple established methods (e.g., CharSwap, synonym replacement, back-translation) to increase coverage of linguistic noise.

Confounding Factors in CoT Prompting: The performance difference between CoT and No-CoT conditions could be influenced by factors beyond the presence of reasoning, such as the increased prompt length or the specific trigger phrase used (“Let’s think step by step”). We mitigated this by keeping non-reasoning content constant across conditions where possible and using a standard, widely-adopted CoT trigger.


\subsection{External Validity} 
External validity concerns the generalizability of our findings to other settings.

Model Selection: Our study focuses on a set of popular open-source code models (7B–13B parameters). The behavior of larger proprietary models (e.g., GPT-4, Claude) or models with different training data may differ. However, the consistent patterns we observe across three distinct model families strengthen the generalizability of the core phenomenon.

Dataset Scope: We evaluate on two benchmarks: MHPP (algorithmic) and BigCodeBench (instructional). While they are representative and widely used, our findings may not extend to other domains like data science scripts, configuration code, or low-level system programming. The observed task-dependence of CoT benefits suggests this is a pertinent threat.

Perturbation Types: Our study is limited to linguistic perturbations at character, word, and sentence levels. Other relevant dimensions, such as perturbations to API names, code examples in the prompt, or program-specific semantics, are not explored. Future work should investigate these.

\subsection{Construct Validity}
Construct validity concerns whether our measurements accurately capture the underlying theoretical constructs.

Evaluation via Execution: We rely on functional correctness (pass@k) as the primary metric. While this is standard, it may not capture subtle semantic errors or reasoning quality issues. A manual inspection of a sample of outputs did not reveal systematic biases, but this remains a limitation.

Measuring Robustness: We define robustness as relative degradation (RD) in pass@k. This captures average performance loss but may obscure changes in the distribution of failures (e.g., a few catastrophic failures vs. many minor ones). Our complementary analysis of failure transitions (Pass→Fail, Fail→Pass) helps mitigate this.

Measuring Uncertainty: We use token-level entropy and probability differential as proxies for the model’s internal uncertainty. While standard, these may not perfectly align with the model’s true confidence, especially in models using speculative decoding or other advanced sampling techniques. In addition, token-level uncertainty signals are not uniformly accessible across all models and inference APIs. Some model deployments do not expose token probabilities or intermediate distributions, which may limit the applicability and reproducibility of uncertainty-based analyses in certain settings.

Defining Structural Anchors: The three anchors we identify are inferred from qualitative analysis of uncertainty traces and failure cases. While they provide a compelling explanatory framework, their precise definition and boundaries are interpretive. More formal methods (e.g., causal intervention) could be used in future work to solidify these constructs.

\subsection{Conclusion Validity}
Conclusion validity concerns the reliability and statistical significance of the inferences drawn.

Statistical Power: We report results based on the full test sets of MHPP (210 problems) and BigCodeBench (251 problems). For RQ2, each perturbation is applied to each prompt, yielding thousands of evaluations per model. This provides high statistical power for detecting main effects. We employ non-parametric Wilcoxon signed-rank tests for significance testing where applicable.

Multiple Comparisons: Across four RQs, multiple models, datasets, temperatures, and perturbation types, we perform numerous statistical comparisons. While we focus on effect sizes and consistent patterns, some individual comparisons may be susceptible to Type I error. We mitigate this by emphasizing large, consistent trends across settings rather than isolated significant results.

By acknowledging these threats, we aim to provide a balanced interpretation of our findings and clarify the boundaries within which our conclusions hold.

%% file: Conclusions.tex
\section{Conclusions} \label{sec:conclusions}

This work shows that the benefits of CoT are conditional rather than universal. Its effectiveness varies across models and datasets, and its apparent gains can weaken substantially under prompt perturbations. We also find that early uncertainty signals carry diagnostic value for downstream failure, and that perturbations reshape the reasoning process itself rather than only degrading the final generated code. These findings argue for viewing CoT not as a monolithic capability, but as a structured process with identifiable points of fragility. The concepts of structural anchors and trajectory deformation provide a useful lens for understanding CoT’s contingent benefits, its characteristic failure modes, and the role of generation-time signals in revealing reasoning instability. Moving forward, the goal should evolve from simply eliciting step-by-step reasoning to building models and prompting strategies that can maintain reasoning integrity under real-world noise and ambiguity.

%% file: Acknowledgement.tex
\section*{Acknowledgement}
This work was supported by: Fonds de Recherche du Québec (FRQ), the Canadian Institute for Advanced Research (CIFAR) as well as the DEEL project CRDPJ 537462-18 funded by the Natural Sciences and Engineering Research Council of Canada (NSERC) and the Consortium for Research and Innovation in Aerospace in Québec (CRIAQ), together with its industrial partners Thales Canada inc, Bell Textron Canada Limited, CAE inc and Bombardier inc.